\documentclass[preprint,aps,prd,superscriptaddress,nofootinbib,tightenlines]{revtex4}

\usepackage{amsfonts}
\usepackage{multirow}
\usepackage{mathrsfs}
\usepackage{graphicx}
\usepackage{amsmath}
\usepackage{amssymb}
\usepackage{bm}
\usepackage{bbm}
\usepackage{color}
\usepackage{array}
\usepackage{diagbox}
\usepackage{float}
\usepackage{makecell}
\usepackage{slashed}
\allowdisplaybreaks[4]

\def\OMIT#1{}

\def\hlinew#1{%
  \noalign{\ifnum0=`}\fi\hrule \@height #1 \futurelet
   \reserved@a\@xhline}
\usepackage{array}
\newcommand{\PreserveBackslash}[1]{\let\temp=\\#1\let\\=\temp}
\newcolumntype{C}[1]{>{\PreserveBackslash\centering}p{#1}}
\newcolumntype{R}[1]{>{\PreserveBackslash\raggedleft}p{#1}}
\newcolumntype{L}[1]{>{\PreserveBackslash\raggedright}p{#1}}

\newcommand{\beq}{\begin{equation}}
\newcommand{\eeq}{\end{equation}}
\newcommand{\bqa}{\begin{eqnarray}}
\newcommand{\eqa}{\end{eqnarray}}

\newcommand\fverb{\setbox\fverbbox=\hbox\bgroup\verb}
\newcommand\fverbdo{\egroup\medskip\noindent%
			\fbox{\unhbox\fverbbox}\ }
\newcommand\fverbit{\egroup\item[\fbox{\unhbox\fverbbox}]}
\newbox\fverbbox

\makeatletter

\newcommand{\Rmnum}[1]{\expandafter\@slowromancap\romannumeral #1@}
\makeatother

\begin{document}
\title{\mbox{}\\[10pt]
Next-to-leading-order QCD corrections to the radiative decays
$\chi_{bJ}\to J/\psi+\gamma$}

\author{Yu-Dong Zhang~\footnote{ydzhang@mails.ccnu.edu.cn}}
\affiliation{Jingchu University of Technology, Jingmen, Hubei 448000, China\vspace{0.2cm}}

\author{Zai-Fu Jiang~\footnote{200707008@jcut.edu.cn}}
\affiliation{Jingchu University of Technology, Jingmen, Hubei 448000, China\vspace{0.2cm}}
\begin{abstract}
	Within the framework of nonrelativistic QCD (NRQCD) factorization,
	we calculate the next-to-leading-order (NLO) QCD corrections to
	$\chi_{bJ}\to J/\psi+\gamma$ with $J=0,1,2$. We determine all six
	independent helicity amplitudes, including the one-loop corrections
	to the QED amplitude and the two-loop corrections to the QCD
	amplitude, and retain the interference between the QED and QCD
	amplitudes in the decay widths. The NLO corrections substantially
	enhance the $\chi_{b0}$ and $\chi_{b1}$ decay widths, while reducing
	the $\chi_{b2}$ decay width by approximately $36\%$. Including the
	QED amplitude leads to sizable destructive interference for
	$\chi_{b0}$ and an enhancement for $\chi_{b2}$, but has only a
	small effect on the $\chi_{b1}$ decay width. Two treatments of
	photon fragmentation, based on the NRQCD matrix element and the
	experimentally determined $J/\psi$ decay constant, yield NLO unpolarized decay widths
	that agree at the few percent level or better.
	Using the decay constant scheme at $\mu_R=m_b$, we obtain NLO central
	branching fractions of $1.03\times10^{-6}$, $1.45\times10^{-5}$,
	and $1.99\times10^{-6}$ for the three channels, respectively.
	The NLO corrections reduce the renormalization scale dependence,
	particularly for $\chi_{b2}$, although a sizable dependence
	remains for $\chi_{b0}$. At fixed $m_b$, the branching fractions decrease with increasing
	$r=m_c/m_b$, more strongly for $\chi_{b1}$ and $\chi_{b2}$
	than for $\chi_{b0}$. For $pp$ collisions at $\sqrt{s}=14\,{\rm TeV}$ and an integrated
	luminosity of $100\,{\rm fb}^{-1}$, we estimate approximately
	$1.9\times10^4$ and $4.8\times10^4$ $\ell^+\ell^-\gamma$ events
	from $\chi_{b0}$ and $\chi_{b2}$ decays, respectively.
	Measurements with the existing Belle $\Upsilon(2S)$ sample
	remain challenging because of the small expected yields. A substantially larger $\Upsilon(2S)$
	sample at Belle II would improve the prospects for observing
	these radiative decays.
\end{abstract}
\maketitle

\section{Introduction\label{sec-introduction}}

Exclusive bottomonium decays into charmonia provide a useful setting for
investigating the interplay between short-distance annihilation and
heavy-quark bound-state dynamics. The bottom- and charm-quark masses
provide perturbative scales, whereas quarkonium formation involves
nonperturbative QCD effects. Within the nonrelativistic QCD (NRQCD)
factorization framework~\cite{Bodwin:1994jh}, the decay amplitude can be
factorized into perturbatively calculable short-distance coefficients
(SDCs) and nonperturbative long-distance matrix elements (LDMEs). 
The SDCs are calculated perturbatively in $\alpha_s$, while
relativistic corrections are organized through the NRQCD
velocity expansion.

On the experimental side, the Belle Collaboration has searched for
charmonium and charmonium-like states in the radiative decays of
$\Upsilon(1S)$ and $\Upsilon(2S)$
~\cite{Belle:2010sgx,Belle:2011jbs}. Belle subsequently observed
$\Upsilon(1S)\to\chi_{c1}+\gamma$ and measured its branching fraction
~\cite{Belle:2019ybg}. Moreover, Belle has also searched for double-charmonium
decays of the spin-triplet $P$-wave bottomonia produced through
$\Upsilon(2S)\to\chi_{bJ}+\gamma$. No significant signals were found in the double-charmonium
final states, and upper limits were set on the corresponding
$\chi_{bJ}$ decay branching fractions~\cite{Shen:2012ei}. At the LHC, the production
of $\chi_b$ states has been studied by LHCb through the radiative
transitions $\chi_b\to\Upsilon(nS)+\gamma$~\cite{LHCb:2014ngh}. 

At leading order (LO), radiative bottomonium decays into charmonia were
studied within NRQCD by Gao \emph{et al.}~\cite{Gao:2007fv}. They
considered $\Upsilon\to\eta_c(\chi_{cJ})+\gamma$ and
$\eta_b(\chi_{bJ})\to J/\psi+\gamma$, including both the QED and QCD
contributions. The decay $\eta_b\to J/\psi+\gamma$ was also
investigated independently in Ref.~\cite{Hao:2006nf}. Analytic LO
results for the radiative decays of vector bottomonia into $S$- and
$P$-wave charmonia were subsequently presented in
Ref.~\cite{Shen:2020woq}. 

Recent advances in multiloop techniques have made it possible to
calculate higher-order QCD corrections to a variety of quarkonium
production and decay processes. At $B$ factories, two-loop QCD
corrections have been obtained for
$e^+e^-\to J/\psi+\eta_c$~\cite{Feng:2019zmt,Huang:2022dfw} and
$e^+e^-\to J/\psi+\chi_{cJ}$~\cite{Sang:2022kub}. In bottomonium decays, the next-to-leading-order (NLO) QCD corrections have been calculated for
$\Upsilon\to\eta_c(\chi_{cJ})+\gamma$~\cite{Zhang:2021ted} and
$\Upsilon\to J/\psi+\eta_c(\chi_{cJ})$~\cite{Zhang:2022nuf}. In
particular, the predicted branching fraction for
$\Upsilon(1S)\to J/\psi+\chi_{c1}$ agrees with the Belle measurement
~\cite{Zhang:2022nuf,Belle:2014wam}. The next-to-next-to-leading-order (NNLO) QCD corrections to the
$C$-even bottomonium decays
$\eta_b(\chi_{bJ})\to J/\psi+J/\psi$ have also been evaluated
~\cite{Zhang:2023mky}. More recently, the NLO QCD corrections to
$\eta_b\to J/\psi+\gamma$ were calculated in
Ref.~\cite{Yang:2026kvh}. They were found to enhance its branching
fraction by a factor of about $3.65$ and to reduce the renormalization scale dependence. By contrast, the corresponding corrections to
$\chi_{bJ}\to J/\psi+\gamma$ have not yet been systematically investigated.

For $\chi_{bJ}\to J/\psi+\gamma$, the QED and QCD amplitudes
start at different orders in the coupling constants. The QED
amplitude starts at tree level and is of order $\alpha^{3/2}$, corresponding to the photon fragmentation process
$\chi_{bJ}\to\gamma+\gamma^*$ followed by
$\gamma^*\to J/\psi$. The pure QCD amplitude first appears at one loop
and is of order $\alpha^{1/2}\alpha_s^2$. The virtual photon propagator
can enhance the photon fragmentation contribution, rendering the QED
amplitude and its interference with the pure QCD amplitude numerically
important~\cite{Gao:2007fv,Hao:2006nf}. Moreover, since the LO QCD decay width starts at order $\alpha\alpha_s^4$, it may exhibit
a strong dependence on the renormalization scale. A complete
higher-order calculation therefore requires both the one-loop QCD
corrections to the QED amplitude and the two-loop corrections to the
pure QCD amplitude.

In this work, we calculate the NLO QCD corrections to
$\chi_{bJ}\to J/\psi+\gamma$ with $J=0,1,2$, at leading order in
the heavy-quark velocity expansion. We determine the six independent
helicity amplitudes and obtain the polarized and unpolarized decay
widths as well as the corresponding branching fractions, retaining
the interference between the QED and QCD amplitudes. For the
photon fragmentation contribution, we consider both the conventional
NRQCD treatment and an alternative scheme in which the
$\gamma^*\to J/\psi$ transition is described by the $J/\psi$ decay
constant extracted from its measured leptonic decay width
~\cite{Zhang:2021ted,Sang:2023liy}. In the latter scheme, the
perturbative matching correction associated with the same transition
is subtracted to avoid double counting.

From an experimental perspective,
$\chi_{bJ}\to J/\psi+\gamma$ requires the reconstruction of
only one $J/\psi$, whose leptonic decays provide clean
experimental signatures. Compared with the double-$J/\psi$
final state, this avoids the additional suppression associated
with requiring a second leptonic $J/\psi$ decay. At $B$ factories, the $\chi_{bJ}$
states can be produced through the E1 transitions
$\Upsilon(2S)\to\chi_{bJ}+\gamma$, a production mechanism already used in Belle
searches~\cite{Shen:2012ei}. A large $\Upsilon(2S)$ data sample collected at Belle II
would therefore provide an opportunity to search for
these radiative decays. At
hadron colliders, the established production of $\chi_{bJ}$ states~\cite{LHCb:2014ngh},
together with the reconstruction of $J/\psi\to\ell\bar\ell$, offers a
complementary search strategy. The predictions presented in this work can provide reliable theoretical
input for future measurements of these decay channels.

The remainder of this paper is organized as follows. In
Sec.~\ref{sec-general-formula}, we present the general formulas for the
helicity amplitudes and the polarized and unpolarized decay widths.
Section~\ref{sec-nrqcd} introduces the NRQCD factorization formula and
the perturbative matching procedure. In
Sec.~\ref{sec-sdc-calculation}, we describe the calculation of the SDCs
at NLO in $\alpha_s$. Section~\ref{sec-phen} is devoted to the
numerical results and phenomenological discussion. The final section
contains our summary.

\section{(Un)polarized decay widths\label{sec-general-formula}}

It is convenient to apply the helicity amplitude formalism~\cite{Haber:1994pe,Jacob:1959at}
to analyze the exclusive decay $\chi_{bJ}\to J/\psi+\gamma$, where $J=0,1,2$.
The magnetic quantum number of the decaying particle with respect to the $z$-axis is denoted by $S_z$, while $\theta$ is the polar angle between the $z$-axis and the momentum of the outgoing $J/\psi$. Let $\lambda_1$ and $\lambda_2$
represent the helicities of the $J/\psi$ and the photon, respectively. Thus,
$\lambda_1=0,\pm1$ and $\lambda_2=\pm1$. The differential decay width is given by
\begin{equation}\label{eq-gen-rate-helicity-jpsigamma}
	\frac{d\Gamma\left[\chi_{bJ}(S_z)\to J/\psi(\lambda_1)\gamma(\lambda_2)\right]}
	{d\cos\theta}
	=\frac{|\mathbf{P}|}{16\pi m_{\chi_{bJ}}^2}
	\left|d^J_{S_z,\lambda_1-\lambda_2}(\theta)\right|^2
	\left|A^{\chi_{bJ}}_{\lambda_1,\lambda_2}\right|^2,
\end{equation}
where $m_{\chi_{bJ}}$ represents the mass of $\chi_{bJ}$, and $\mathbf{P}$ denotes
the three-momentum of the $J/\psi$ in the $\chi_{bJ}$ rest frame. Its magnitude is
\begin{equation}\label{eq-jpsigamma-momentum}
	|\mathbf{P}|=
	\frac{\lambda^{1/2}\left(m_{\chi_{bJ}}^2,m_{J/\psi}^2,0\right)}
	{2m_{\chi_{bJ}}}
	=\frac{m_{\chi_{bJ}}^2-m_{J/\psi}^2}{2m_{\chi_{bJ}}},
\end{equation}
where the K\"all\'en function is defined via
$\lambda(x,y,z)=x^2+y^2+z^2-2xy-2xz-2yz$. Angular momentum conservation requires $|\lambda_1-\lambda_2|\leq J$.
The angular distribution is dictated by the Wigner function
$d^J_{S_z,\lambda_1-\lambda_2}(\theta)$, while
$A^{\chi_{bJ}}_{\lambda_1,\lambda_2}$ is the helicity amplitude that contains
all the nontrivial dynamics. In our convention, the helicity amplitude is defined in
the reference configuration in which the $J/\psi$ moves along the positive $z$-axis,
so that $S_z=\lambda_1-\lambda_2$.

Since the $J/\psi$ and the photon are distinguishable particles, the integration is performed over the full angular range
$-1\leq\cos\theta\leq1$, and no identical particle symmetry factor
is required. Averaging over all possible $\chi_{bJ}$ polarizations, one obtains
\begin{equation}\label{eq-gen-rate-helicity-int-jpsigamma}
	\begin{aligned}
		\Gamma\left[\chi_{bJ}\to J/\psi(\lambda_1)\gamma(\lambda_2)\right]
		&=\frac{|\mathbf{P}|}{16\pi m_{\chi_{bJ}}^2}
		\left|A^{\chi_{bJ}}_{\lambda_1,\lambda_2}\right|^2
		\int_{-1}^{1}\frac{1}{2J+1}\sum_{S_z=-J}^{J}
		\left|d^J_{S_z,\lambda_1-\lambda_2}(\theta)\right|^2d\cos\theta
		\\
		&=\frac{|\mathbf{P}|}{8\pi(2J+1)m_{\chi_{bJ}}^2}
		\left|A^{\chi_{bJ}}_{\lambda_1,\lambda_2}\right|^2.
	\end{aligned}
\end{equation}

The helicity amplitudes are not independent. Due to parity invariance~\cite{Haber:1994pe},
they satisfy
\begin{equation}\label{eq-helicity-parity-jpsigamma}
	A^{\chi_{bJ}}_{\lambda_1,\lambda_2}
	=(-1)^J A^{\chi_{bJ}}_{-\lambda_1,-\lambda_2}.
\end{equation}
Explicitly, the parity relations among the allowed helicity amplitudes are
\begin{subequations}\label{eq-helicity-parity-explicit-jpsigamma}
	\begin{align}
		A^{\chi_{b0}}_{1,1}
		&=A^{\chi_{b0}}_{-1,-1},
		\\
		A^{\chi_{b1}}_{0,1}
		&=-A^{\chi_{b1}}_{0,-1},
		&
		A^{\chi_{b1}}_{1,1}
		&=-A^{\chi_{b1}}_{-1,-1},
		\\
		A^{\chi_{b2}}_{0,1}
		&=A^{\chi_{b2}}_{0,-1},
		&
		A^{\chi_{b2}}_{1,1}
		&=A^{\chi_{b2}}_{-1,-1},
		&
		A^{\chi_{b2}}_{1,-1}
		&=A^{\chi_{b2}}_{-1,1}.
	\end{align}
\end{subequations}
There is no additional interchange relation because the two final-state particles are
not identical. Thus, there is one independent helicity amplitude for $\chi_{b0}$,
two for $\chi_{b1}$, and three for $\chi_{b2}$. We choose
$A^{\chi_{b0}}_{1,1}$,
$A^{\chi_{b1}}_{1,1}$, $A^{\chi_{b1}}_{0,1}$, $A^{\chi_{b2}}_{1,-1}$, $A^{\chi_{b2}}_{1,1}$, and
$A^{\chi_{b2}}_{0,1}$ as the independent helicity amplitudes.

It is straightforward to obtain the unpolarized decay widths for
$\chi_{bJ}\to J/\psi+\gamma$ by summing over all the allowed helicity channels:
\begin{subequations}\label{eq-gen-rate-helicity-explicit-jpsigamma}
	\begin{align}
		\Gamma(\chi_{b0}\to J/\psi+\gamma)
		&=\frac{|\mathbf{P}|}{8\pi m_{\chi_{b0}}^2}
		\left(2\left|A^{\chi_{b0}}_{1,1}\right|^2\right),
		\\
		\Gamma(\chi_{b1}\to J/\psi+\gamma)
		&=\frac{|\mathbf{P}|}{24\pi m_{\chi_{b1}}^2}
		\left(2\left|A^{\chi_{b1}}_{1,1}\right|^2
		+2\left|A^{\chi_{b1}}_{0,1}\right|^2\right),
		\\
		\Gamma(\chi_{b2}\to J/\psi+\gamma)
		&=\frac{|\mathbf{P}|}{40\pi m_{\chi_{b2}}^2}
		\left(2\left|A^{\chi_{b2}}_{1,-1}\right|^2
		+2\left|A^{\chi_{b2}}_{1,1}\right|^2
		+2\left|A^{\chi_{b2}}_{0,1}\right|^2\right).
	\end{align}
\end{subequations}
\section{NRQCD factorization for the helicity amplitudes\label{sec-nrqcd}}

At the lowest order in the velocity expansion, the $\chi_{bJ}$ and
$J/\psi$ can be identified with the leading color-singlet
$b\bar b({}^3P_J)$ and $c\bar c({}^3S_1)$ Fock states, respectively.
According to NRQCD factorization~\cite{Bodwin:1994jh}, the helicity
amplitude for $\chi_{bJ}\to J/\psi+\gamma$ can be factorized as
\begin{equation}\label{eq-nrqcd-jpsigamma}
	A^{\chi_{bJ}}_{\lambda_1,\lambda_2}
	=\sqrt{2m_{\chi_{bJ}}}\sqrt{2m_{J/\psi}}\,
	c^{\chi_{bJ}}_{\lambda_1,\lambda_2}
	\frac{\sqrt{\langle\mathcal O\rangle_{\chi_{bJ}}}}{m_b^{5/2}}
	\frac{\sqrt{\langle\mathcal O\rangle_{J/\psi}}}{m_c^{3/2}}.
\end{equation}
Here $\lambda_1$ and $\lambda_2$ denote the helicities of the $J/\psi$
and photon, respectively, and
$c^{\chi_{bJ}}_{\lambda_1,\lambda_2}$ denotes the dimensionless SDC.
The powers of $m_b$ and $m_c$ in
(\ref{eq-nrqcd-jpsigamma}) follow from the mass dimensions of the $P$-wave
and $S$-wave NRQCD matrix elements.

The nonperturbative LDMEs are defined by
$\langle\mathcal O\rangle_H\equiv
|\langle0|\mathcal O_H|H\rangle|^2$, with
\begin{subequations}\label{eq-nrqcd-bilinears-jpsigamma}
	\begin{align}
		\mathcal O_{\chi_{b0}}
		&=\frac{1}{\sqrt{3}}\chi_b^\dagger
		\left(-\frac{i}{2}\overleftrightarrow{\mathbf D}\cdot
		\boldsymbol{\sigma}\right)\psi_b,
		\\
		\mathcal O_{\chi_{b1}}
		&=\frac{1}{\sqrt{2}}\chi_b^\dagger
		\left(-\frac{i}{2}\overleftrightarrow{\mathbf D}\times
		\boldsymbol{\sigma}\right)\cdot
		\boldsymbol{\epsilon}_{\chi_{b1}}\psi_b,
		\\
		\mathcal O_{\chi_{b2}}
		&=\chi_b^\dagger
		\left(-\frac{i}{2}\overleftrightarrow D^{(i}\sigma^{j)}\right)
		\epsilon_{\chi_{b2}}^{ij}\psi_b,
		\\
		\mathcal O_{J/\psi}
		&=\chi_c^\dagger\boldsymbol{\sigma}\cdot
		\boldsymbol{\epsilon}_{J/\psi}\psi_c.
	\end{align}
\end{subequations}
Here $\psi_Q$ and $\chi_Q^\dagger$ denote the Pauli spinor fields
annihilating a heavy quark and antiquark, respectively,
while $\boldsymbol{\epsilon}_{\chi_{b1}}$,
$\epsilon_{\chi_{b2}}^{ij}$, and
$\boldsymbol{\epsilon}_{J/\psi}$ represent the corresponding
polarization vectors and tensor.

The factor $\sqrt{2m_{\chi_{bJ}}}\sqrt{2m_{J/\psi}}$ on the right-hand
side of (\ref{eq-nrqcd-jpsigamma}) arises because relativistic
normalization is adopted for the quarkonium states in the helicity
amplitude, whereas conventional nonrelativistic normalization is used for
the LDMEs. Since relativistic corrections are not considered in this work,
we set $m_{\chi_{bJ}}=2m_b$ and $m_{J/\psi}=2m_c$ in the perturbative
matching. The physical quarkonium masses are retained in the state-normalization
factors.

The SDCs are determined by perturbative matching.
 We replace the physical quarkonia by free on-shell heavy-quark
pairs $b\bar b({}^3P_J^{[1]})$ and
$c\bar c({}^3S_1^{[1]})$, respectively, and compute the
amplitude in both full QCD and NRQCD. With relativistic
normalization for the free heavy-quark states, the matching relation reads
\begin{equation}\label{eq-partonic-matching-jpsigamma}
	\begin{aligned}
		\mathcal A^{(J)}_{\lambda_1,\lambda_2}
		&=
		c^{\chi_{bJ}}_{\lambda_1,\lambda_2}
		\frac{2m_b\sqrt{2N_c}}{m_b^{5/2}}
		\frac{2m_c\sqrt{2N_c}}{m_c^{3/2}}
		\\
		&=
		c^{\chi_{bJ}}_{\lambda_1,\lambda_2}
		\frac{8N_c}{m_b^{3/2}m_c^{1/2}}.
	\end{aligned}
\end{equation}
Here $\mathcal A^{(J)}_{\lambda_1,\lambda_2}$ denotes the amplitude at the quark level after the $P$-wave
projection. The factor $2m_b\sqrt{2N_c}$ is obtained after
extracting the term linear in the initial relative momentum
and applying the normalized spin-orbital projector.
Solving for the SDC gives
\begin{equation}\label{eq-sdc-extraction-jpsigamma}
	c^{\chi_{bJ}}_{\lambda_1,\lambda_2}
	=\mathcal A^{(J)}_{\lambda_1,\lambda_2}
	\frac{m_b^{3/2}m_c^{1/2}}{8N_c}.
\end{equation}

We next specify the kinematics and projectors used in the matching. The
momenta of the heavy quarks are assigned as
\begin{equation}\label{eq-partonic-momenta-jpsigamma}
	p_b=\frac{Q}{2}+q,\qquad
	\bar p_b=\frac{Q}{2}-q,\qquad
	p_c=\bar p_c=\frac{P}{2},\qquad Q=P+k,
\end{equation}
where $q$ is the relative momentum of the initial $b\bar b$ pair and $k$
is the photon momentum. The on-shell conditions imply
\begin{equation}\label{eq-partonic-onshell-jpsigamma}
Q\cdot q=0,\qquad	Q^2=4E_b^2,\qquad
	P^2=4m_c^2,\qquad k^2=0.
\end{equation}
Here $E_b=\sqrt{m_b^2+\boldsymbol q^{\,2}}$ is the heavy-quark
energy in the rest frame of the initial $b\bar b$ pair. After the amplitude is expanded through first order in $q$,
we set $q=0$, so that $E_b=m_b$ and $Q^2=4m_b^2$. We treat the $J/\psi$ at leading order in the velocity expansion
throughout the calculation and therefore set the relative
momentum of the final $c\bar c$ pair to zero.

The covariant spin and color projectors for the
initial $b\bar b$ and final $c\bar c$ pairs in
color-singlet, spin-triplet states are
~\cite{Petrelli:1997ge,Bodwin:2002cfe}
\begin{subequations}\label{eq-spin-projectors-jpsigamma}
	\begin{align}
		\overline{\Pi}_{1,b}^{\mu}(Q,q)
		={}&\frac{1}{4\sqrt{2}E_b(E_b+m_b)}
		(\slashed p_b+m_b)(\slashed Q+2E_b)\gamma^\mu
		(\slashed{\bar p}_b-m_b)
		\otimes\frac{\mathbf 1_c}{\sqrt{N_c}},
		\label{eq-spin-projector-bb-jpsigamma}
		\\
		\Pi_{1,c}^{\nu}(P)
		={}&-\frac{1}{8\sqrt{2}m_c^2}
		\left(\frac{\slashed P}{2}-m_c\right)\gamma^\nu
		(\slashed P+2m_c)
		\left(\frac{\slashed P}{2}+m_c\right)
		\otimes\frac{\mathbf 1_c}{\sqrt{N_c}}.
		\label{eq-spin-projector-cc-jpsigamma}
	\end{align}
\end{subequations}
The first projector retains the dependence on the relative momentum
$q$ required for the $P$-wave projection.

Let $\mathcal M_\alpha(q)$ denote the amputated amplitude at the quark level for
$b\bar b\to c\bar c+\gamma$, and let
$\epsilon_{\chi_{bJ}}^{\mu\rho}$ denote the polarization tensor that
couples the spin and orbital angular momenta of the initial $b\bar b$ pair
to total angular momentum $J$. The amplitude at the quark level for definite
external polarizations is obtained as
\begin{equation}\label{eq-pwave-projection-jpsigamma}
	\begin{aligned}
		\mathcal A^{(J)}_{\lambda_1,\lambda_2}
		={}&\epsilon_{\chi_{bJ}}^{\mu\rho}
		\epsilon_{J/\psi}^{*\nu}(\lambda_1)
		\epsilon_\gamma^{*\alpha}(\lambda_2)
		\frac{\partial}{\partial q^\rho}
		\operatorname{Tr}\!\left[
		\overline\Pi_{1,b,\mu}(Q,q)\,
		\mathcal M_\alpha(q)\,
		\Pi_{1,c,\nu}(P)
		\right]|_{q=0}.
	\end{aligned}
\end{equation}
The $q$ derivative acts on both the initial-state projector and the hard amplitude, while the final $c\bar c({}^3S_1)$ state is taken at leading order in $v$.

For later convenience, we separate the total SDC into its QED and QCD
parts:
\begin{equation}\label{eq-sdc-decomposition-jpsigamma}
	c^{\chi_{bJ}}_{\lambda_1,\lambda_2}
	=d^{\chi_{bJ}}_{\lambda_1,\lambda_2}
	+f^{\chi_{bJ}}_{\lambda_1,\lambda_2}.
\end{equation}
Here $d^{\chi_{bJ}}_{\lambda_1,\lambda_2}$ and
$f^{\chi_{bJ}}_{\lambda_1,\lambda_2}$ denote the QED and QCD SDCs,
respectively. They are obtained by applying the matching relation
(\ref{eq-sdc-extraction-jpsigamma}) separately to the QED and QCD parts of the amplitude.

The QED contribution starts with the photon fragmentation process
$\chi_{bJ}\to\gamma+\gamma^*$ followed by $\gamma^*\to J/\psi$, and is of
order $\alpha^{3/2}$ at leading order. The QCD contribution starts at order
$\alpha^{1/2}\alpha_s^2$ and is generated by the one-loop amplitude. The two SDCs can be expanded in $\alpha_s$ as
\begin{subequations}\label{eq-sdcs-expansion-jpsigamma}
	\begin{align}
		d^{\chi_{bJ}}_{\lambda_1,\lambda_2}
		={}&\alpha^{3/2}\left[
		d^{\chi_{bJ},(0)}_{\lambda_1,\lambda_2}
		+\frac{\alpha_s(\mu_R)}{\pi}
		d^{\chi_{bJ},(1)}_{\lambda_1,\lambda_2}
		+\mathcal O(\alpha_s^2)
		\right],
		\\
		f^{\chi_{bJ}}_{\lambda_1,\lambda_2}
		={}&\alpha^{1/2}\,\alpha_s^2(\mu_R)\left\{
		f^{\chi_{bJ},(0)}_{\lambda_1,\lambda_2}
		+\frac{\alpha_s(\mu_R)}{\pi}\left[
		\frac{\beta_0}{2}\ln\frac{\mu_R^2}{m_b^2}
		f^{\chi_{bJ},(0)}_{\lambda_1,\lambda_2}
		+f^{\chi_{bJ},(1)}_{\lambda_1,\lambda_2}
		\right]
		+\mathcal O(\alpha_s^2)
		\right\}.
	\end{align}
\end{subequations}
Here
$\beta_0=\frac{11}{3}C_A-\frac{4}{3}T_f n_f$, with
$n_f=n_l+n_h$ and $n_h=n_c+n_b$. The coefficient of the logarithmic term in
(\ref{eq-sdcs-expansion-jpsigamma}) is fixed by renormalization group invariance.

For the photon fragmentation contribution, one may alternatively replace
the perturbative NRQCD treatment of $\gamma^*\to J/\psi$ by the $J/\psi$
decay constant extracted from its leptonic decay width. We define
\begin{equation}\label{eq-jpsi-decay-constant}
	\langle0|\bar c\gamma^\mu c|J/\psi(P,\lambda)\rangle
	=f_{J/\psi}m_{J/\psi}\epsilon_{J/\psi}^\mu(\lambda),
	\qquad
	\Gamma(J/\psi\to e^+e^-)
	=\frac{4\pi\alpha^2e_c^2}{3m_{J/\psi}}f_{J/\psi}^2,
\end{equation}
where $e_c=2/3$. We refer to the direct NRQCD treatment in
(\ref{eq-nrqcd-jpsigamma}) as the ${\rm QED}_{\rm I}$ scheme. In the
${\rm QED}_{\rm II}$ scheme, the replacement
\begin{equation}\label{eq-qedii-substitution-jpsigamma}
	\sqrt{\langle\mathcal O\rangle_{J/\psi}}
	\longrightarrow\sqrt{m_c}\,f_{J/\psi}
\end{equation}
is made only in the QED amplitude. Accordingly,
\begin{equation}\label{eq-qedii-factorization-jpsigamma}
	\left(A^{\chi_{bJ},{\rm QED}}_{\lambda_1,\lambda_2}\right)_{{\rm QED}_{\rm II}}
	=\sqrt{2m_{\chi_{bJ}}}\sqrt{2m_{J/\psi}}\,
	d^{\chi_{bJ}}_{\lambda_1,\lambda_2}
	\frac{\sqrt{\langle\mathcal O\rangle_{\chi_{bJ}}}}{m_b^{5/2}}
	\frac{f_{J/\psi}}{m_c}.
\end{equation}
Since the radiative corrections associated with
$\gamma^*\to J/\psi$ are already encoded in the experimentally
determined decay constant $f_{J/\psi}$, they must be subtracted
from the QED SDC to avoid double counting. Using~\cite{Beneke:1997jm,Czarnecki:1997vz,Marquard:2014pea}
\begin{equation}\label{eq-vector-current-matching-jpsigamma}
	f_{J/\psi}
	=\sqrt{\frac{\langle\mathcal O\rangle_{J/\psi}}{m_c}}
	\left[1-2C_F\frac{\alpha_s}{\pi}
	+\mathcal O(\alpha_s^2)\right],
\end{equation}
one obtains
\begin{equation}\label{eq-qed-scheme-conversion-jpsigamma}
	\left(d^{\chi_{bJ},(1)}_{\lambda_1,\lambda_2}\right)_{{\rm QED}_{\rm II}}
	=\left(d^{\chi_{bJ},(1)}_{\lambda_1,\lambda_2}\right)_{{\rm QED}_{\rm I}}
	+2C_Fd^{\chi_{bJ},(0)}_{\lambda_1,\lambda_2}.
\end{equation}
The QCD coefficient $f^{\chi_{bJ}}_{\lambda_1,\lambda_2}$ is unchanged
between the two schemes.

The NLO decay widths in both QED schemes are obtained by squaring of the corresponding total helicity amplitudes,
	with the QED and QCD amplitudes truncated at NLO in $\alpha_s$~\cite{Jia:2024ini,Huang:2023pmn,Zeng:2026ois}.

The LO QCD SDCs exhibit different asymptotic behaviors
for different helicity configurations in the limit
$r=m_c/m_b\to0$. Their leading asymptotic expressions
are presented in Sec.~\ref{sec-sdc-calculation}.

\section{Perturbative calculation 
	\label{sec-sdc-calculation}}

\begin{figure}
	\centering
	\includegraphics[width=0.92\textwidth]
	{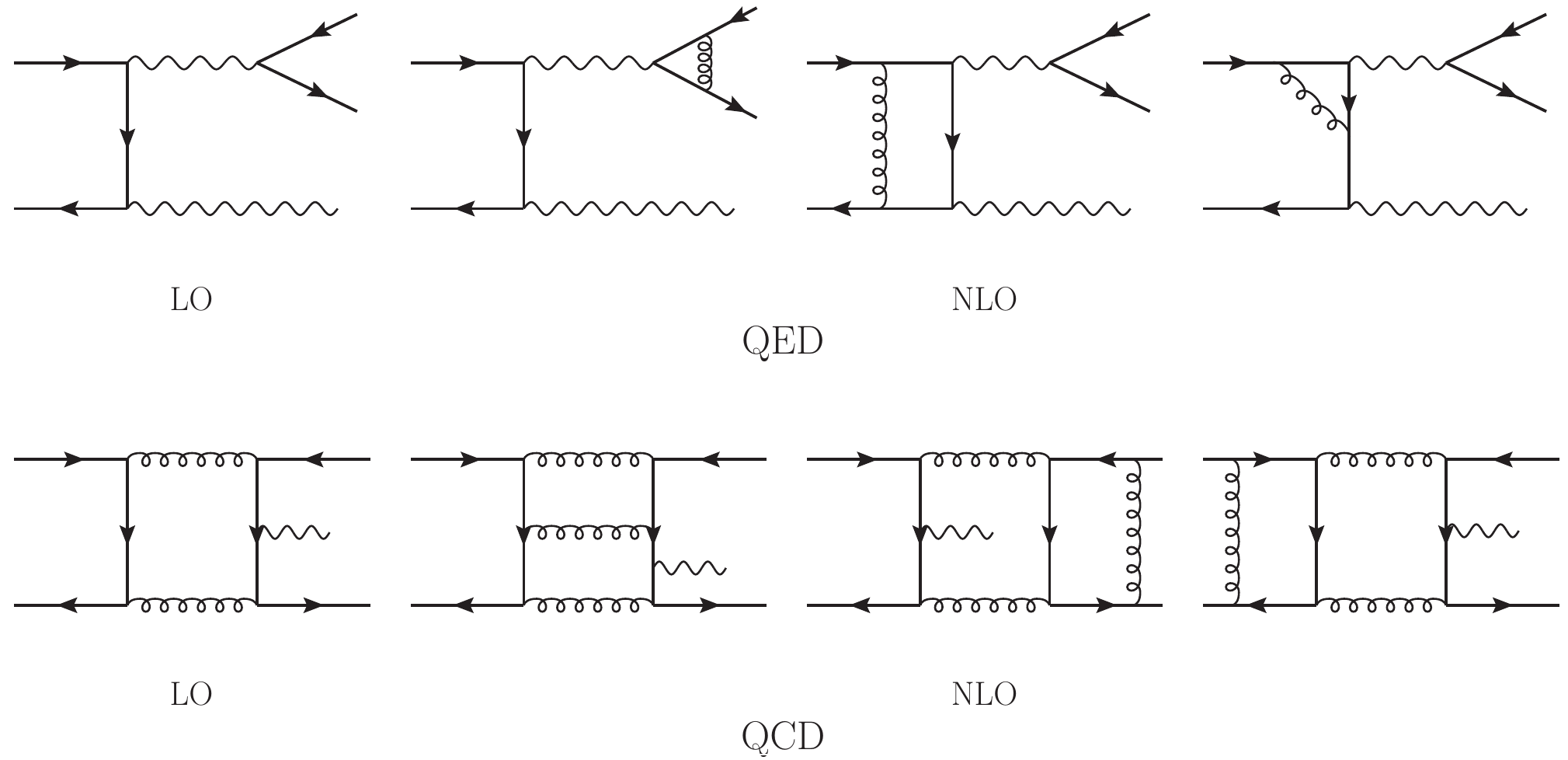}
	\caption{Representative Feynman diagrams for
		$\chi_{bJ}\to J/\psi+\gamma$, drawn with
		\texttt{JaxoDraw}~\cite{Binosi:2003yf}.}
	\label{fig-feynman-diagrams-jpsigamma}
\end{figure}

At the quark level, the perturbative matching amounts to calculating
the amplitude for
$b\bar b({}^3P_J^{[1]})\to c\bar c({}^3S_1^{[1]})+\gamma$.
We separate the amplitudes into QED and QCD contributions
according to their coupling structure. The former starts with the tree-level
photon fragmentation process
$\chi_{bJ}\to\gamma+\gamma^*$ followed by
$\gamma^*\to J/\psi$, and is of order $\alpha^{3/2}$. The latter first appears at
one loop and is of order $\alpha^{1/2}\alpha_s^2$. Therefore, to obtain both
contributions through relative order $\alpha_s$, one has to calculate the
one-loop correction to the QED amplitude and the two-loop correction to
the pure QCD amplitude. Some
representative diagrams are shown in
Fig.~\ref{fig-feynman-diagrams-jpsigamma}.

We employ \texttt{FeynArts}~\cite{Hahn:2000kx} to generate the Feynman
diagrams and the corresponding amplitudes at the quark level. The covariant
spin and color projectors given in Sec.~\ref{sec-nrqcd} are then used to
project the $b\bar b$ and $c\bar c$ pairs onto the color-singlet
states. The $P$-wave projection is carried out by differentiating the
projected amplitude with respect to the relative momentum $q$ of the
initial $b\bar b$ pair and subsequently setting $q$ to zero. In contrast,
the final $c\bar c({}^3S_1)$ pair is kept at zero relative momentum from the outset. The Dirac and color algebra is performed with
\texttt{FeynCalc}~\cite{Shtabovenko:2016sxi} and
\texttt{FormLink}~\cite{Feng:2012tk}. We obtain the six independent helicity amplitudes specified in Sec.~\ref{sec-general-formula} by contracting the projected amplitude with the polarization tensors and vectors.

In terms of the mass ratio $r=m_c/m_b$, the LO QED SDCs read
\begin{subequations}\label{eq-lo-qed-sdcs-jpsigamma}
	\begin{align}
		d^{\chi_{b0},(0)}_{1,1}
		&=\frac{4\sqrt{3}\pi^{3/2}(r^2-3)}
		{81\sqrt{r}(r^2-1)},
		\label{eq-lo-qed-sdc-chib0-jpsigamma}
		\\
		d^{\chi_{b1},(0)}_{1,1}
		&=\frac{4\sqrt{2}\,\pi^{3/2}r^{3/2}}
		{27(r^2-1)},
		&
		d^{\chi_{b1},(0)}_{0,1}
		&=\frac{4\sqrt{2}\,\pi^{3/2}\sqrt{r}}
		{27(1-r^2)},
		\label{eq-lo-qed-sdc-chib1-jpsigamma}
		\\
		d^{\chi_{b2},(0)}_{1,1}
		&=\frac{4\sqrt{6}\,\pi^{3/2}r^{3/2}}
		{81(r^2-1)},
		&
		d^{\chi_{b2},(0)}_{1,-1}
		&=\frac{8\pi^{3/2}}
		{27\sqrt{r}(r^2-1)},
		&
		d^{\chi_{b2},(0)}_{0,1}
		&=\frac{4\sqrt{2}\,\pi^{3/2}\sqrt{r}}
		{27(r^2-1)}.
		\label{eq-lo-qed-sdc-chib2-jpsigamma}
	\end{align}
\end{subequations}
Here $e_b=-1/3$ and $e_c=2/3$ have been included
in the reduced SDCs.

Beyond tree level, we expand the integrands in the relative momentum
$q$ prior to loop integration. Within the method of regions, this
amounts to retaining the hard region that determines the SDCs~\cite{Beneke:1997zp}. There are 10 one-loop QED diagrams and 6
one-loop QCD diagrams. We use \texttt{Apart}~\cite{Feng:2012iq} for partial-fraction
decomposition and \texttt{FIRE}~\cite{Smirnov:2014hma} for
integration-by-parts (IBP) reduction. The reduction yields 14 one-loop master integrals (MIs)
and 552 two-loop MIs. The one-loop MIs are
evaluated analytically with
\texttt{Package-X}~\cite{Patel:2015tea}, while the two-loop MIs are
evaluated numerically using the auxiliary-mass-flow method~\cite{Liu:2017jxz,Liu:2022mfb,Liu:2021wks,Liu:2022chg}.

Both ultraviolet (UV) and infrared (IR) divergences are regularized in $D=4-2\epsilon$
dimensions. The heavy-quark masses and fields are renormalized in the
on-shell scheme, while $\alpha_s$ is renormalized in the
$\overline{\rm MS}$ scheme. Since the pure QCD amplitude starts at
one loop, only the one-loop renormalization constants $Z_m$, $Z_2$,
and $Z_{\alpha_s}$ are required at the present accuracy and must be expanded to the required order in $\epsilon$. The renormalized SDCs are free of both UV and IR poles.

Since the exact analytic expressions for
$f^{\chi_{bJ},(0)}_{\lambda_1,\lambda_2}$ are rather lengthy, we
present only their leading asymptotic forms in the limit $r\to0$.
Factoring out the leading power of $r$, we write
\begin{equation}\label{eq-nlo-qcd-sdc-asymptotic-jpsigamma}
	\left.
	f^{\chi_{bJ},(0)}_{\lambda_1,\lambda_2}
	\right|_{\mathrm{asy}}
	=
	\frac{4\sqrt{\pi}}{81}\,
	r^{1/2+|\lambda_1|}
	g^{\chi_{bJ},(0)}_{\lambda_1,\lambda_2}.
\end{equation}
The explicit
power of $r$ has been factored out to make the helicity scaling
manifest. The coefficients
$g^{\chi_{bJ},(0)}_{\lambda_1,\lambda_2}$ contain no additional
power suppression in $r$, although they may contain logarithms of
$r$. They read
\begin{subequations}\label{eq-nlo-qcd-asymptotic-coefficients-jpsigamma}
	\begin{align}
		g^{\chi_{b0},(0)}_{1,1}
		={}&2\sqrt{3}\Big[
		(6\ln2-1)\ln(-r^2+i\epsilon)
		+\ln^2 2
		+(1-8i\pi)\ln2
		-\pi^2-3
		\Big],
		\label{eq-nlo-qcd-asymptotic-chib0-jpsigamma}
		\\
		g^{\chi_{b1},(0)}_{1,1}
		={}&6\sqrt{2}\Big[
		\ln(-r^2+i\epsilon)
		-6\ln^2 2
		+(7+12i\pi)\ln2
		-8i\pi
		\Big],
		\\
		g^{\chi_{b1},(0)}_{0,1}
		={}&12\sqrt{2}\Big[
		-\ln^2 2
		+(2+2i\pi)\ln2
		-1-2i\pi
		\Big],
		\label{eq-nlo-qcd-asymptotic-chib1-jpsigamma}
		\\
		g^{\chi_{b2},(0)}_{1,1}
		={}&-2\sqrt{6}\Big[
		\ln(-r^2+i\epsilon)
		+26\ln^2 2
		-(37+52i\pi)\ln2
		+15+36i\pi
		\Big],
		\\
		g^{\chi_{b2},(0)}_{1,-1}
		={}&-4\Big[
		3\ln^2(-r^2+i\epsilon)
		+12\ln^2 2
		-(14+24i\pi)\ln2
		+\pi^2+25i\pi+8
		\Big],
		\\
		g^{\chi_{b2},(0)}_{0,1}
		={}&6\sqrt{2}\Big[
		6\ln^2 2
		-(8+12i\pi)\ln2
		+3+8i\pi
		\Big].
		\label{eq-nlo-qcd-asymptotic-chib2-jpsigamma}
	\end{align}
\end{subequations}

For consistency with Refs.~\cite{Gao:2007fv,Zhang:2023mky,Yang:2026kvh}, we take the heavy-quark masses to be
$m_b=4.7\,{\rm GeV}$ and $m_c=1.5\,{\rm GeV}$.
For the QED contribution, $d^{(0)}_{\lambda_1,\lambda_2}$ is common to the
${\rm QED}_{\rm I}$ and ${\rm QED}_{\rm II}$ schemes, while the
corresponding NLO coefficients are related by
Eq.~(\ref{eq-qed-scheme-conversion-jpsigamma}). For the pure QCD
contribution, $f^{(0)}_{\lambda_1,\lambda_2}$ and  $f^{(1)}_{\lambda_1,\lambda_2}$ denote the LO
and NLO coefficients, respectively. The fermion-loop contributions to
$f^{(1)}_{\lambda_1,\lambda_2}$ are decomposed according to quark flavor.
Here $n_l$ denotes the number of massless light-quark flavors, while
$n_c$ and $n_b$ denote the numbers of charm- and bottom-quark flavors,
respectively. We take $n_l=3$, $n_c=1$, and $n_b=1$. The dependence on
$n_l$, $n_c$, and $n_b$ is kept explicit in the two-loop SDCs. The
resulting QED and pure QCD helicity SDCs are summarized in
Table~\ref{tab-sdcs-jpsigamma}.
\begin{table*}[htbp]
	\centering
	\scriptsize
	\setlength{\tabcolsep}{3.5pt}
	\renewcommand{\arraystretch}{1.45}
	\caption{
		Dimensionless QED and QCD helicity SDCs for
		$\chi_{bJ}\to J/\psi+\gamma$.
		The flavor dependence of the fermion-loop part of
		$f^{(1)}_{\lambda_1,\lambda_2}$ is kept explicit.
	}
	\label{tab-sdcs-jpsigamma}
	\resizebox{\textwidth}{!}{%
		\begin{tabular}{|c|c|c|c|c|c|c|}
			\hline
			\multicolumn{7}{|l|}{
				$m_b=4.7\,{\rm GeV},\quad
				m_c=1.5\,{\rm GeV}
				\ ;\qquad \mathrm{SDCs}$
			}
			\\
			\hline
			
			\multirow{2}{*}{$H$}
			& \multirow{2}{*}{$(\lambda_1,\lambda_2)$}
			& \multicolumn{3}{c|}{QED}
			& \multicolumn{2}{c|}{QCD}
			\\
			\cline{3-7}
			
			&
			& $d^{(0)}_{\lambda_1,\lambda_2}$
			& $\bigl(d^{(1)}_{\lambda_1,\lambda_2}\bigr)_{{\rm QED}_{\rm I}}$
			& $\bigl(d^{(1)}_{\lambda_1,\lambda_2}\bigr)_{{\rm QED}_{\rm II}}$
			& $f^{(0)}_{\lambda_1,\lambda_2}$
			& $f^{(1)}_{\lambda_1,\lambda_2}$
			\\
			\hline
			
			$\chi_{b0}$ & $(1,1)$
			& $2.720$
			& $-7.053$
			& $0.201$
			& $-1.088-0.179i$
			&
			$\begin{aligned}
				&-3.051-9.447i+(-0.419+0.793i)n_l\\[-2pt]
				&\quad+(-0.388+0.638i)n_c+(-0.762-0.025i)n_b
			\end{aligned}$
			\\
			\hline
			
			\multirow{2}{*}{$\chi_{b1}$}
			& $(0,1)$
			& $0.734$
			& $-2.568$
			& $-0.611$
			& $-0.186-1.553i$
			&
			$\begin{aligned}
				&-2.252-3.196i+(-1.275-0.471i)n_l\\[-2pt]
				&\quad+(-0.922-0.629i)n_c+(-0.461-1.099i)n_b
			\end{aligned}$
			\\
			\cline{2-7}
			
			& $(1,1)$
			& $-0.234$
			& $0.840$
			& $0.216$
			& $-0.075+0.530i$
			&
			$\begin{aligned}
				&1.442+0.475i+(0.448+0.230i)n_l\\[-2pt]
				&\quad+(0.278+0.278i)n_c+(0.065+0.376i)n_b
			\end{aligned}$
			\\
			\hline
			
			\multirow{3}{*}{$\chi_{b2}$}
			& $(0,1)$
			& $-0.734$
			& $3.936$
			& $1.979$
			& $-0.132+0.140i$
			&
			$\begin{aligned}
				&-2.460-2.118i+(0.117+0.166i)n_l\\[-2pt]
				&\quad+(0.134+0.073i)n_c+(-0.094+0.099i)n_b
			\end{aligned}$
			\\
			\cline{2-7}
			
			& $(1,1)$
			& $-0.135$
			& $1.564$
			& $1.204$
			& $0.002-0.040i$
			&
			$\begin{aligned}
				&-0.989-0.217i+(-0.047+0.021i)n_l\\[-2pt]
				&\quad+(0.002-0.012i)n_c+(-0.008-0.024i)n_b
			\end{aligned}$
			\\
			\cline{2-7}
			
			& $(1,-1)$
			& $-3.252$
			& $16.977$
			& $8.305$
			& $-0.470+1.725i$
			&
			$\begin{aligned}
				&-3.958-8.799i+(1.597+0.619i)n_l\\[-2pt]
				&\quad+(1.134+0.421i)n_c+(-0.221+1.136i)n_b
			\end{aligned}$
			\\
			\hline
		\end{tabular}%
	}
\end{table*}

\section{Phenomenology and discussion\label{sec-phen}}

We now use the helicity amplitudes obtained in
Secs.~\ref{sec-nrqcd} and \ref{sec-sdc-calculation} to investigate the
radiative decays
$\chi_{bJ}\to J/\psi+\gamma$.
In the perturbative amplitudes, the quarkonium masses are identified
with $2m_b$ and $2m_c$, as specified in Sec.~III, consistently with
the on-shell kinematics used in the matching.  For the physical quarkonium masses entering the normalization
factors,  we take~\cite{ParticleDataGroup:2024cfk}
\begin{equation}\label{eq-physical-masses-phen-jpsigamma}
	\begin{aligned}
		m_{\chi_{b0}}&=9.85944~{\rm GeV}, &
		m_{\chi_{b1}}&=9.89278~{\rm GeV},\\
		m_{\chi_{b2}}&=9.91221~{\rm GeV}, &
		m_{J/\psi}&=3.09690~{\rm GeV},
	\end{aligned}
\end{equation}
and the heavy-quark pole masses to be
$m_b=4.7~{\rm GeV}$ and $m_c=1.5~{\rm GeV}$~\cite{Gao:2007fv,Zhang:2023mky,Yang:2026kvh}. The strong coupling
$\alpha_s(\mu_R)$ is evolved at two-loop accuracy using \texttt{RunDec}~\cite{Chetyrkin:2000yt}. We use a fixed QED coupling, $\alpha=1/131$, throughout our numerical analysis. This value has also been used in studies of bottomonium radiative decays~\cite{Zhang:2021ted} and charmonium production~\cite{Sang:2009jc,Fan:2012dy}. We set the central renormalization scale to $\mu_R=m_b$ and estimate the scale uncertainty by varying $\mu_R$ between $2m_c$ and $2m_b$.

The NRQCD LDMEs for the $S$-wave and $P$-wave quarkonia
are estimated from the radial wave functions at the origin
and their first derivatives, respectively:
\begin{subequations}\label{eq-ldme-inputs-phen-jpsigamma}
	\begin{align}
		\langle\mathcal O\rangle_{J/\psi}
		&\simeq
		\frac{N_c}{2\pi}
		\left|R^{c\bar c}_{1S}(0)\right|^2
		=\frac{N_c}{2\pi}\times0.810~{\rm GeV}^3,
		\label{eq-jpsi-ldme-input-phen-jpsigamma}
		\\
		\langle\mathcal O\rangle_{\chi_{bJ}}
		&\simeq
		\frac{3N_c}{2\pi}
		\left|R_{1P}^{\prime\,b\bar b}(0)\right|^2
		=\frac{3N_c}{2\pi}\times1.417~{\rm GeV}^5.
		\label{eq-chib-ldme-input-phen-jpsigamma}
	\end{align}
\end{subequations}
Here $N_c=3$. The radial wave functions are taken from
Ref.~\cite{Eichten:1995ch}, where they were evaluated using the
Buchm\"uller--Tye potential model.
	The choice of potential model introduces uncertainties in the
	theoretical predictions.
	For example, the four models considered in
	Ref.~\cite{Eichten:1995ch} give values of
	$|R_{1S}^{c\bar c}(0)|^2$ between $0.810$ and
	$1.454\,\mathrm{GeV}^3$.
	For the lowest $P$-wave bottomonium states, estimates of
	$|R_{1P}^{\prime b\bar b}(0)|^2$ range from $0.93$ to
	$2.07\,\mathrm{GeV}^5$~\cite{Chung:2021efj}.
	These differences reflect both the choice of potential and the
	heavy-quark masses used in the models.
	With all other inputs fixed, the decay widths are proportional to
	$\langle O\rangle_{\chi_{bJ}}$. In the QED$_{\mathrm{II}}$ scheme,
	$\langle O\rangle_{J/\psi}$ still enters the pure QCD contribution
	and the interference between the QCD and QED amplitudes.
	At leading order in the velocity
expansion, heavy-quark spin symmetry implies a common LDME for the three
$\chi_{bJ}$ states.

In the ${\rm QED}_{\rm II}$ scheme, the $J/\psi$ LDME in the QED
amplitude is replaced by the decay constant defined in
Eq.~(\ref{eq-jpsi-decay-constant}).  Numerically, we use~\cite{ParticleDataGroup:2024cfk}
\begin{equation}\label{eq-fjpsi-input-phen-jpsigamma}
	\begin{aligned}
		\Gamma(J/\psi\to e^+e^-)
		&=5.52915\times10^{-6}~{\rm GeV},\\
		f_{J/\psi}
		&=0.39729~{\rm GeV}.
	\end{aligned}
\end{equation}
The predictions in Table~\ref{tab-widths-jpsigamma}
are evaluated in the ${\rm QED}_{\rm II}$ scheme.
A comparison with the ${\rm QED}_{\rm I}$ scheme is
presented in Table~\ref{tab-qed-scheme-comparison-jpsigamma}. 
We adopt the full decay widths estimated from the E1 transitions
$\chi_{bJ}\to\gamma\Upsilon(1S)$ in Ref.~\cite{Zhang:2023mky}:
\begin{equation}
	\begin{aligned}
		\Gamma_{\chi_{b0}}
		= 1.144^{+0.185}_{-0.140}\,{\rm MeV},\qquad
		\Gamma_{\chi_{b1}}
		= 0.079^{+0.005}_{-0.004}\,{\rm MeV},\qquad
		\Gamma_{\chi_{b2}}
		= 0.176^{+0.010}_{-0.009}\,{\rm MeV}.
	\end{aligned}
	\label{eq-chibj-full-widths}
\end{equation}

As shown in Table~\ref{tab-widths-jpsigamma}, the
contributions from different helicity configurations
differ considerably. For the $\chi_{b1}$ decay channel,
the helicity $(0,1)$ contribution accounts for approximately $89\%$ and $91\%$ of the
unpolarized decay width at LO and NLO, respectively.
This contribution is much larger than that
from helicity $(1,1)$.
For the $\chi_{b2}$ decay channel, helicity $(1,-1)$ accounts for approximately
$98\%$ and $95\%$ of the unpolarized decay width at
LO and NLO, respectively.  The NLO corrections are sizable for all three decay channels, enhancing
the decay widths of $\chi_{b0}$ and $\chi_{b1}$, while reducing the $\chi_{b2}$ decay
width by approximately $36\%$. Additionally, the $\chi_{b0}$
and $\chi_{b1}$ decay widths are comparable at NLO,
approximately $1.18~{\rm eV}$ and $1.14~{\rm eV}$,
respectively. The much larger $\chi_{b1}$ branching
fraction therefore mainly reflects its smaller full
decay width. Despite the negative NLO correction,
the $\chi_{b2}$ branching fraction remains larger than
that of $\chi_{b0}$ because $\chi_{b2}$ has a much
smaller full decay width. 

\begin{table*}[htbp]
	\centering
	\small
	\setlength{\tabcolsep}{4pt}
	\renewcommand{\arraystretch}{1.35}
	\caption{
		NRQCD predictions for the helicity decay widths and branching
		fractions of $\chi_{bJ}\to J/\psi+\gamma$ at LO and NLO in
		$\alpha_s$, evaluated at $m_b=4.7\,{\rm GeV}$ and
		$m_c=1.5\,{\rm GeV}$.
		$\Gamma_{\rm QCD}$ denotes the pure QCD contribution, while
		$\Gamma_{\rm total}$ includes the QCD contribution, the ${\rm QED}_{\rm II}$
		contribution, and their interference. The branching fractions are obtained by dividing the
		unpolarized decay widths for
		$\chi_{bJ}\to J/\psi+\gamma$ by the corresponding
		$\chi_{bJ}$ full decay widths.
		The first uncertainty originates from varying the renormalization
		scale $\mu_R$ from $2m_c$ to $2m_b$. For the branching fractions,
		the second uncertainty arises from uncertainties in the
		full decay widths of the $\chi_{bJ}$ .
	}
	\label{tab-widths-jpsigamma}
	\resizebox{\textwidth}{!}{%
		\begin{tabular}{|c|c|c|c|c|c|c|c|}
			\hline
			\multicolumn{8}{|l|}{
				$m_b=4.7\,{\rm GeV},\quad
				m_c=1.5\,{\rm GeV}
				\ ;\qquad
				\Gamma\bigl(\times10^{-2}\,{\rm eV}\bigr)$
			}
			\\
			\hline
			
			\multirow{2}{*}{$H$}
			& \multirow{2}{*}{$(\lambda_1,\lambda_2)$}
			& \multicolumn{3}{c|}{LO}
			& \multicolumn{3}{c|}{NLO}
			\\
			\cline{3-8}
			
			&
			& $\Gamma_{\rm QCD}$
			& $\Gamma_{\rm total}$
			& $\operatorname{Br}\bigl(\times10^{-6}\bigr)$
			& $\Gamma_{\rm QCD}$
			& $\Gamma_{\rm total}$
			& $\operatorname{Br}\bigl(\times10^{-6}\bigr)$
			\\
			\hline
			
			$\chi_{b0}$
			& $(1,1)$
			& $43.82^{+37.75}_{-22.54}$
			& $21.00^{+27.86}_{-14.32}$
			& $0.37^{+0.49+0.05}_{-0.25-0.05}$
			& $91.38^{+37.81}_{-32.23}$
			& $59.02^{+32.56}_{-25.82}$
			& $1.03^{+0.57+0.14}_{-0.45-0.14}$
			\\
			\hline
			
			\multirow{2}{*}{$\chi_{b1}$}
			& $(0,1)$
			& $87.91^{+75.73}_{-45.22}$
			& $86.98^{+75.27}_{-44.84}$
			& \multirow{2}{*}{
				$8.22^{+7.10+0.44}_{-4.23-0.49}$
			}
			& $159.96^{+54.55}_{-52.10}$
			& $155.81^{+52.70}_{-50.66}$
			& \multirow{2}{*}{
				$14.49^{+4.87+0.77}_{-4.68-0.86}$
			}
			\\
			\cline{2-4}\cline{6-7}
			
			& $(1,1)$
			& $10.28^{+8.86}_{-5.29}$
			& $10.47^{+8.91}_{-5.34}$
			&
			& $16.14^{+5.23}_{-4.95}$
			& $15.89^{+4.98}_{-4.78}$
			&
			\\
			\hline
			
			\multirow{3}{*}{$\chi_{b2}$}
			& $(0,1)$
			& $1.33^{+1.15}_{-0.69}$
			& $2.53^{+1.47}_{-0.95}$
			& \multirow{3}{*}{
				$3.12^{+2.40+0.19}_{-1.46-0.17}$
			}
			& $2.78^{+1.84}_{-1.12}$
			& $4.48^{+2.20}_{-1.45}$
			& \multirow{3}{*}{
				$1.99^{+0.08+0.12}_{-0.53-0.11}$
			}
			\\
			\cline{2-4}\cline{6-7}
			
			& $(1,1)$
			& $0.06^{+0.05}_{-0.03}$
			& $0.07^{+0.05}_{-0.03}$
			&
			& $0.31^{+0.36}_{-0.17}$
			& $0.35^{+0.37}_{-0.18}$
			&
			\\
			\cline{2-4}\cline{6-7}
			
			& $(1,-1)$
			& $114.71^{+98.81}_{-59.00}$
			& $134.78^{+103.89}_{-63.22}$
			&
			& $70.13^{+3.57}_{-23.38}$
			& $82.83^{+4.21}_{-25.82}$
			&
			\\
			\hline
		\end{tabular}%
	}
\end{table*}
\begin{figure}[H]
	\centering
	\includegraphics[width=0.48\textwidth]{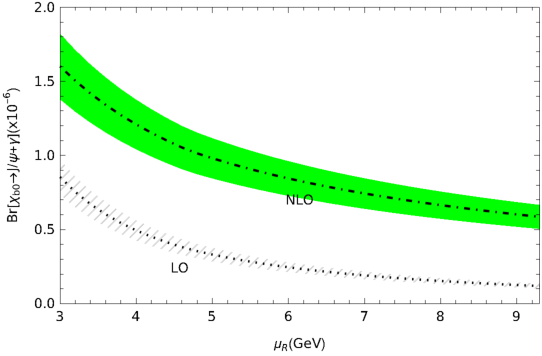}
	\hfill
	\includegraphics[width=0.48\textwidth]{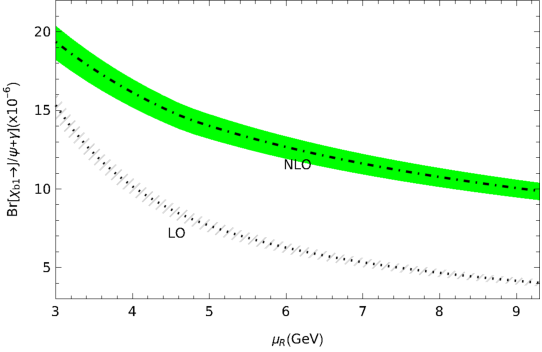}
	\medskip
	\includegraphics[width=0.48\textwidth]{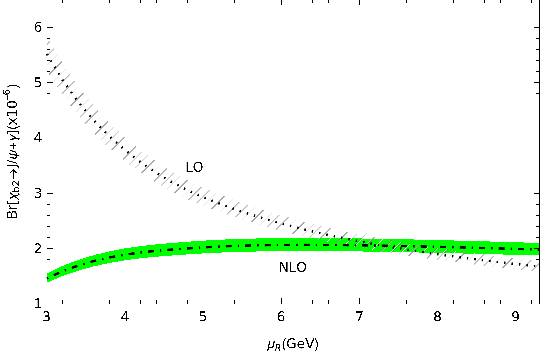}
	\caption{Branching fractions for
		$\chi_{bJ}\to J/\psi+\gamma$ as functions of the
		renormalization scale $\mu_R$. The gray hatched and green
		bands represent the LO and NLO predictions, respectively,
		with the bands reflecting the uncertainties in the full decay widths of the $\chi_{bJ}$.}
	\label{fig-scale-dependence-jpsigamma}
\end{figure}

The impact of the QED contribution can be seen by comparing
$\Gamma_{\rm total}$ with $\Gamma_{\rm QCD}$. For $\chi_{b0}$, including
the QED amplitude reduces the pure QCD decay width by approximately
$52\%$ at LO and $35\%$ at NLO, reflecting sizable destructive
interference between the QED and QCD amplitudes. For $\chi_{b1}$, the net QED effect remains small at both
LO and NLO. By contrast,
for $\chi_{b2}$, the QED and interference contributions together increase the pure QCD result by approximately $18\%$ at LO and $20\%$
at NLO.

Fig.~\ref{fig-scale-dependence-jpsigamma} shows the dependence of
the branching fractions on $\mu_R$. The gray hatched and green bands
represent the LO and NLO predictions, respectively, with the uncertainties in the full decay widths of the
$\chi_{bJ}$. The $\mu_R$-dependence is substantially reduced at NLO for
$\chi_{b2}$, with a more moderate reduction for $\chi_{b1}$.
A sizable $\mu_R$-dependence persists for $\chi_{b0}$, indicating
that higher-order corrections may still be important for this channel.

\begin{table*}[htbp]
	\centering
	\small
	\setlength{\tabcolsep}{4pt}
	\renewcommand{\arraystretch}{1.35}
	\caption{
		Comparison of the QED and QCD+QED helicity contributions for
		$\chi_{bJ}\to J/\psi+\gamma$ in the two QED schemes.
		$\Gamma_{{\rm QED}_{\rm I}}$ and
		$\Gamma_{{\rm QED}_{\rm II}}$ denote the QED contributions
		evaluated in the ${\rm QED}_{\rm I}$ and ${\rm QED}_{\rm II}$
		schemes, respectively.
		The combined widths
		$\Gamma_{{\rm QCD}+{\rm QED}_{\rm I}}$ and
		$\Gamma_{{\rm QCD}+{\rm QED}_{\rm II}}$ include the QCD
		contribution, the corresponding QED contribution, and their
		interference.
		The uncertainties are estimated by varying the renormalization
		scale $\mu_R$ from $2m_c$ to $2m_b$.
	}
	\label{tab-qed-scheme-comparison-jpsigamma}
	\resizebox{\textwidth}{!}{%
		\begin{tabular}{|c|c|c|c|c|c|c|c|c|c|}
			\hline
			\multicolumn{10}{|l|}{
				$m_b=4.7\,{\rm GeV},\quad
				m_c=1.5\,{\rm GeV}
				\ ;\qquad
				\Gamma\bigl(\times10^{-2}\,{\rm eV}\bigr)$
			}
			\\
			\hline
			
			\multirow{2}{*}{$H$}
			& \multirow{2}{*}{$(\lambda_1,\lambda_2)$}
			& \multicolumn{4}{c|}{LO}
			& \multicolumn{4}{c|}{NLO}
			\\
			\cline{3-10}
			
			&
			& $\Gamma_{{\rm QED}_{\rm I}}$
			& $\Gamma_{{\rm QED}_{\rm II}}$
			& $\Gamma_{{\rm QCD}+{\rm QED}_{\rm I}}$
			& $\Gamma_{{\rm QCD}+{\rm QED}_{\rm II}}$
			& $\Gamma_{{\rm QED}_{\rm I}}$
			& $\Gamma_{{\rm QED}_{\rm II}}$
			& $\Gamma_{{\rm QCD}+{\rm QED}_{\rm I}}$
			& $\Gamma_{{\rm QCD}+{\rm QED}_{\rm II}}$
			\\
			\hline
			
			$\chi_{b0}$
			& $(1,1)$
			& $7.05$
			& $4.31$
			& $16.19^{+25.12}_{-12.03}$
			& $21.00^{+27.86}_{-14.32}$
			& $4.75^{+0.35}_{-0.34}$
			& $4.36^{+0.01}_{-0.01}$
			& $57.80^{+33.62}_{-26.34}$
			& $59.02^{+32.56}_{-25.82}$
			\\
			\hline
			
			\multirow{2}{*}{$\chi_{b1}$}
			& $(0,1)$
			& $0.51$
			& $0.31$
			& $86.83^{+75.15}_{-44.73}$
			& $86.98^{+75.27}_{-44.84}$
			& $0.29^{+0.03}_{-0.03}$
			& $0.28^{+0.01}_{-0.01}$
			& $155.70^{+52.90}_{-50.72}$
			& $155.81^{+52.70}_{-50.66}$
			\\
			\cline{2-10}
			
			& $(1,1)$
			& $0.05$
			& $0.03$
			& $10.54^{+8.93}_{-5.35}$
			& $10.47^{+8.91}_{-5.34}$
			& $0.0295^{+0.0033}_{-0.0032}$
			& $0.0279^{+0.0006}_{-0.0006}$
			& $15.88^{+4.99}_{-4.78}$
			& $15.89^{+4.98}_{-4.78}$
			\\
			\hline
			
			\multirow{3}{*}{$\chi_{b2}$}
			& $(0,1)$
			& $0.51$
			& $0.31$
			& $2.98^{+1.56}_{-1.03}$
			& $2.53^{+1.47}_{-0.95}$
			& $0.20^{+0.04}_{-0.04}$
			& $0.21^{+0.02}_{-0.02}$
			& $4.46^{+2.06}_{-1.35}$
			& $4.48^{+2.20}_{-1.45}$
			\\
			\cline{2-10}
			
			& $(1,1)$
			& $0.02$
			& $0.01$
			& $0.07^{+0.05}_{-0.03}$
			& $0.07^{+0.05}_{-0.03}$
			& $(7.06^{+12.22}_{-6.26})\times10^{-4}$
			& $(1.58^{+0.94}_{-0.73})\times10^{-3}$
			& $0.33^{+0.35}_{-0.17}$
			& $0.35^{+0.37}_{-0.18}$
			\\
			\cline{2-10}
			
			& $(1,-1)$
			& $10.01$
			& $6.13$
			& $142.54^{+105.30}_{-64.40}$
			& $134.78^{+103.89}_{-63.22}$
			& $4.10^{+0.80}_{-0.74}$
			& $4.16^{+0.30}_{-0.29}$
			& $82.70^{+4.74}_{-26.63}$
			& $82.83^{+4.21}_{-25.82}$
			\\
			\hline
		\end{tabular}%
	}
\end{table*}
Table~\ref{tab-qed-scheme-comparison-jpsigamma}
compares the predictions in the
${\rm QED}_{\rm I}$ and ${\rm QED}_{\rm II}$ schemes. Although the LO QED SDCs are common to
the two schemes, their LO QED decay widths differ because the
${\rm QED}_{\rm I}$ scheme uses the NRQCD LDME, whereas the
${\rm QED}_{\rm II}$ scheme employs the $J/\psi$ decay constant
extracted from its leptonic decay width. For the dominant helicity channels, the LO
QED decay widths in the ${\rm QED}_{\rm II}$ scheme are about $40\%$ smaller
than those in the ${\rm QED}_{\rm I}$ scheme. At NLO, these differences decrease to approximately $10\%$ or less.
Once the pure QCD contribution and the interference between the QED and QCD amplitudes are
included, the scheme dependence is further suppressed. After summing over the helicities, we find that the predictions in the two schemes differ by approximately $30\%$, $0.1\%$, and $6\%$ at LO for $\chi_{b0}$, $\chi_{b1}$, and
$\chi_{b2}$, respectively. At NLO, the corresponding differences
decrease to approximately $2.1\%$, $0.1\%$, and $0.2\%$. This agreement indicates that, for the inputs used here,
the NLO predictions for the
$\chi_{bJ}\to J/\psi+\gamma$ decay widths depend only
weakly on the choice between the
${\rm QED}_{\rm I}$ and ${\rm QED}_{\rm II}$ schemes.

\begin{table*}[htbp]
	\centering
	\small
	\setlength{\tabcolsep}{3pt}
	\renewcommand{\arraystretch}{1.35}
	\caption{
		Branching fractions for $\chi_{bJ}\to J/\psi+\gamma$ at LO and NLO
		as functions of $r=m_c/m_b$, with $m_b=4.7\,{\rm GeV}$ fixed
		and $m_c$ varied from $1.3$ to $1.7\,{\rm GeV}$.
		The uncertainties originate from the renormalization scale variation
		and the full decay widths.
	}
	\label{tab-br-r-dependence-jpsigamma1}
	\resizebox{\textwidth}{!}{%
		\begin{tabular}{|c|c|c|c|c|c|c|c|c|c|c|}
			\hline
			\multicolumn{11}{|l|}{
				$m_b=4.7\,{\rm GeV}\ ;\qquad
				{\rm Br}\bigl(\times10^{-6}\bigr)$
			}
			\\
			\hline
			
			$H$
			& \diagbox{Order}{$r$}
			& $0.277$ & $0.287$ & $0.298$ & $0.309$ & $0.319$
			& $0.330$ & $0.340$ & $0.351$ & $0.362$
			\\
			\hline
			
			\multirow{2}{*}{$\chi_{b0}$}
			& LO
			& $0.37^{+0.65}_{-0.27}$
			& $0.37^{+0.64}_{-0.27}$
			& $0.37^{+0.63}_{-0.26}$
			& $0.37^{+0.62}_{-0.26}$
			& $0.37^{+0.61}_{-0.26}$
			& $0.36^{+0.59}_{-0.26}$
			& $0.36^{+0.58}_{-0.25}$
			& $0.36^{+0.57}_{-0.25}$
			& $0.35^{+0.55}_{-0.24}$
			\\
			\cline{2-11}
			
			& NLO
			& $1.06^{+0.84}_{-0.54}$
			& $1.06^{+0.83}_{-0.53}$
			& $1.05^{+0.82}_{-0.53}$
			& $1.04^{+0.80}_{-0.52}$
			& $1.03^{+0.79}_{-0.51}$
			& $1.02^{+0.78}_{-0.51}$
			& $1.01^{+0.76}_{-0.50}$
			& $0.99^{+0.75}_{-0.49}$
			& $0.98^{+0.74}_{-0.48}$
			\\
			\hline
			
			\multirow{2}{*}{$\chi_{b1}$}
			& LO
			& $11.14^{+10.72}_{-5.84}$
			& $10.29^{+9.90}_{-5.39}$
			& $9.53^{+9.17}_{-4.99}$
			& $8.84^{+8.51}_{-4.64}$
			& $8.22^{+7.92}_{-4.31}$
			& $7.67^{+7.39}_{-4.02}$
			& $7.16^{+6.90}_{-3.76}$
			& $6.70^{+6.46}_{-3.52}$
			& $6.28^{+6.06}_{-3.30}$
			\\
			\cline{2-11}
			
			& NLO
			& $19.94^{+8.25}_{-6.97}$
			& $18.34^{+7.55}_{-6.40}$
			& $16.91^{+6.93}_{-5.89}$
			& $15.63^{+6.38}_{-5.43}$
			& $14.49^{+5.90}_{-5.03}$
			& $13.46^{+5.47}_{-4.67}$
			& $12.54^{+5.08}_{-4.34}$
			& $11.70^{+4.73}_{-4.05}$
			& $10.94^{+4.42}_{-3.78}$
			\\
			\hline
			
			\multirow{2}{*}{$\chi_{b2}$}
			& LO
			& $4.44^{+3.81}_{-2.10}$
			& $4.05^{+3.49}_{-1.92}$
			& $3.71^{+3.21}_{-1.76}$
			& $3.40^{+2.96}_{-1.62}$
			& $3.12^{+2.73}_{-1.49}$
			& $2.87^{+2.52}_{-1.38}$
			& $2.65^{+2.34}_{-1.27}$
			& $2.45^{+2.17}_{-1.18}$
			& $2.27^{+2.01}_{-1.10}$
			\\
			\cline{2-11}
			
			& NLO
			& $2.92^{+0.27}_{-0.83}$
			& $2.64^{+0.25}_{-0.76}$
			& $2.40^{+0.23}_{-0.71}$
			& $2.18^{+0.21}_{-0.65}$
			& $1.99^{+0.20}_{-0.61}$
			& $1.82^{+0.18}_{-0.56}$
			& $1.67^{+0.17}_{-0.53}$
			& $1.53^{+0.16}_{-0.49}$
			& $1.41^{+0.15}_{-0.46}$
			\\
			\hline
		\end{tabular}%
	}
\end{table*}
\begin{table*}[htbp]
	\centering
	\small
	\setlength{\tabcolsep}{3pt}
	\renewcommand{\arraystretch}{1.35}
	\caption{
		Same as Table~\ref{tab-br-r-dependence-jpsigamma1}, but with $m_c=1.5\,{\rm GeV}$ fixed
		and $m_b$ varied from $4.5$ to $4.9\,{\rm GeV}$.
	}
	\label{tab-br-r-dependence-jpsigamma2}
	\resizebox{\textwidth}{!}{%
		\begin{tabular}{|c|c|c|c|c|c|c|c|c|c|c|}
			\hline
			\multicolumn{11}{|l|}{
				$m_c=1.5\,{\rm GeV}\ ;\qquad
				{\rm Br}\bigl(\times10^{-6}\bigr)$
			}
			\\
			\hline
			
			$H$
			& \diagbox{Order}{$r$}
			& $0.306$ & $0.309$ & $0.313$ & $0.316$ & $0.319$
			& $0.323$ & $0.326$ & $0.330$ & $0.333$
			\\
			\hline
			
			\multirow{2}{*}{$\chi_{b0}$}
			& LO
			& $0.27^{+0.46}_{-0.19}$
			& $0.29^{+0.49}_{-0.21}$
			& $0.31^{+0.53}_{-0.22}$
			& $0.34^{+0.56}_{-0.24}$
			& $0.37^{+0.61}_{-0.26}$
			& $0.40^{+0.65}_{-0.28}$
			& $0.43^{+0.70}_{-0.30}$
			& $0.46^{+0.75}_{-0.32}$
			& $0.50^{+0.80}_{-0.35}$
			\\
			\cline{2-11}
			
			& NLO
			& $0.74^{+0.57}_{-0.37}$
			& $0.81^{+0.62}_{-0.40}$
			& $0.88^{+0.67}_{-0.44}$
			& $0.95^{+0.73}_{-0.47}$
			& $1.03^{+0.79}_{-0.51}$
			& $1.12^{+0.86}_{-0.56}$
			& $1.22^{+0.94}_{-0.61}$
			& $1.32^{+1.02}_{-0.66}$
			& $1.44^{+1.11}_{-0.72}$
			\\
			\hline
			
			\multirow{2}{*}{$\chi_{b1}$}
			& LO
			& $6.71^{+6.47}_{-3.52}$
			& $7.06^{+6.80}_{-3.70}$
			& $7.42^{+7.15}_{-3.89}$
			& $7.81^{+7.52}_{-4.10}$
			& $8.22^{+7.92}_{-4.31}$
			& $8.66^{+8.35}_{-4.54}$
			& $9.13^{+8.80}_{-4.79}$
			& $9.63^{+9.28}_{-5.05}$
			& $10.16^{+9.79}_{-5.33}$
			\\
			\cline{2-11}
			
			& NLO
			& $11.49^{+4.51}_{-3.93}$
			& $12.17^{+4.82}_{-4.18}$
			& $12.89^{+5.15}_{-4.44}$
			& $13.66^{+5.51}_{-4.72}$
			& $14.49^{+5.90}_{-5.03}$
			& $15.38^{+6.32}_{-5.36}$
			& $16.33^{+6.77}_{-5.71}$
			& $17.35^{+7.26}_{-6.09}$
			& $18.45^{+7.79}_{-6.50}$
			\\
			\hline
			
			\multirow{2}{*}{$\chi_{b2}$}
			& LO
			& $2.60^{+2.25}_{-1.24}$
			& $2.72^{+2.36}_{-1.29}$
			& $2.85^{+2.48}_{-1.36}$
			& $2.98^{+2.60}_{-1.42}$
			& $3.12^{+2.73}_{-1.49}$
			& $3.27^{+2.87}_{-1.57}$
			& $3.43^{+3.01}_{-1.65}$
			& $3.60^{+3.16}_{-1.73}$
			& $3.78^{+3.33}_{-1.82}$
			\\
			\cline{2-11}
			
			& NLO
			& $1.59^{+0.17}_{-0.52}$
			& $1.68^{+0.18}_{-0.54}$
			& $1.78^{+0.19}_{-0.56}$
			& $1.88^{+0.19}_{-0.58}$
			& $1.99^{+0.20}_{-0.61}$
			& $2.11^{+0.21}_{-0.63}$
			& $2.24^{+0.21}_{-0.66}$
			& $2.37^{+0.22}_{-0.68}$
			& $2.51^{+0.23}_{-0.71}$
			\\
			\hline
		\end{tabular}%
	}
\end{table*}
\begin{figure}[H]
	\centering
	\includegraphics[width=0.32\textwidth]{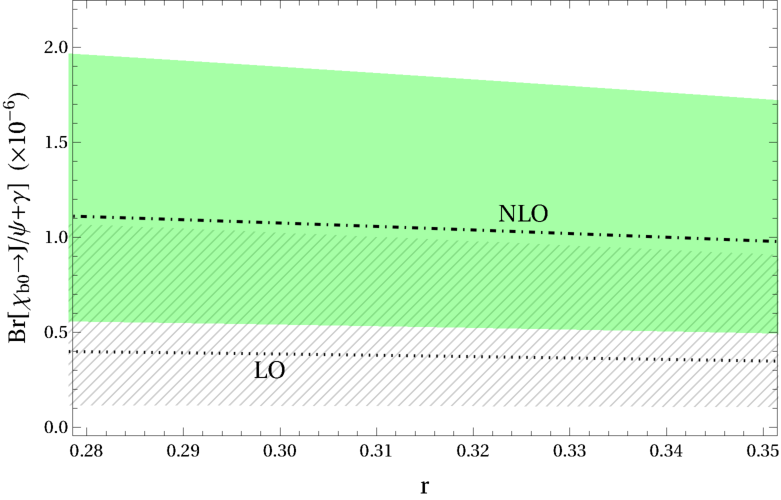}
	\includegraphics[width=0.32\textwidth]{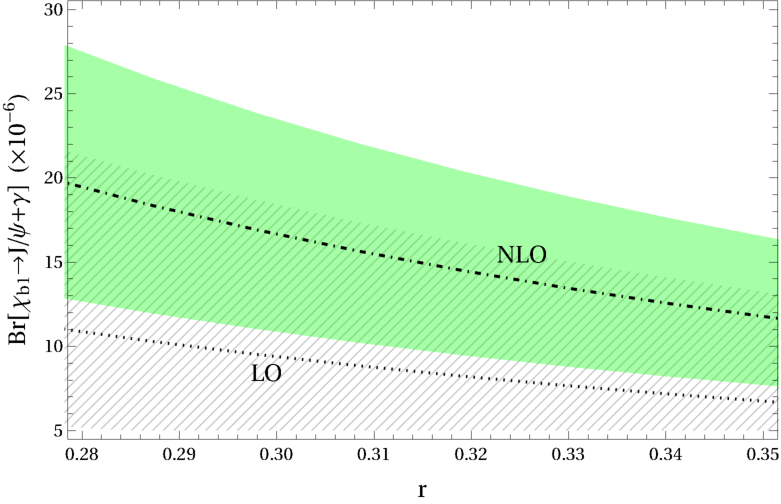}
	\includegraphics[width=0.32\textwidth]{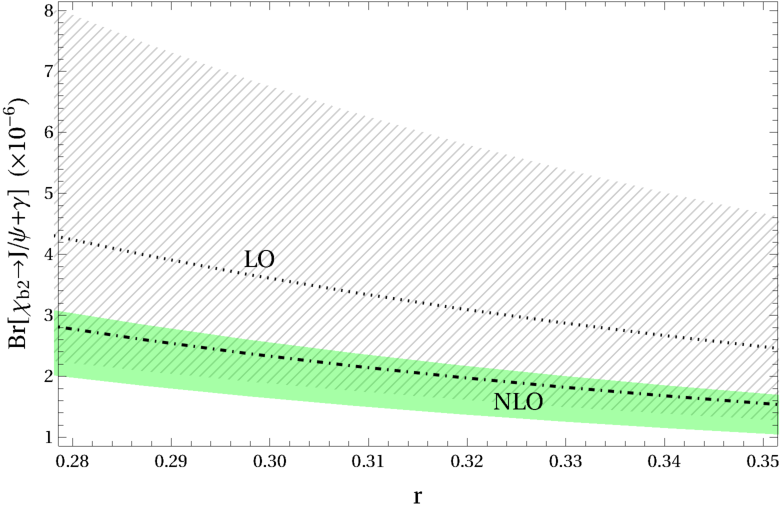}
	\includegraphics[width=0.32\textwidth]{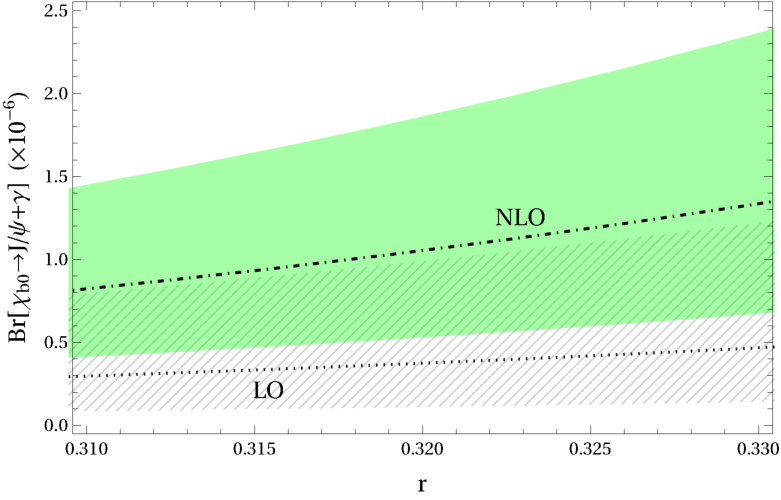}
	\includegraphics[width=0.32\textwidth]{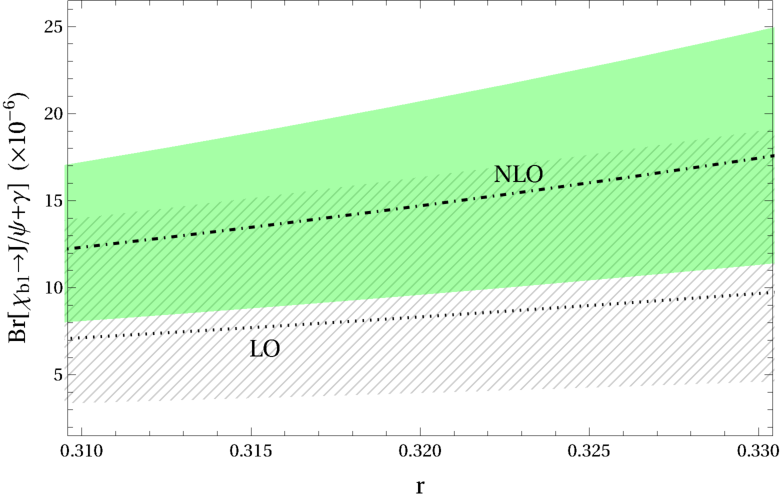}
	\includegraphics[width=0.32\textwidth]{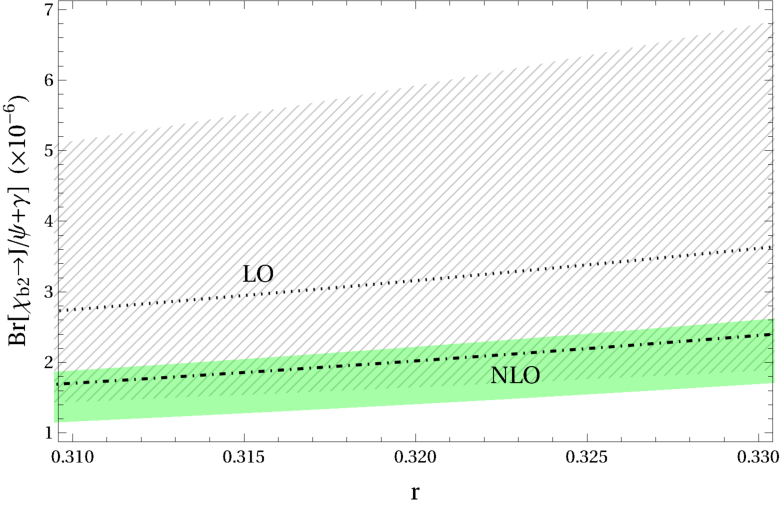}
	\caption{Branching fractions for $\chi_{bJ}\to J/\psi+\gamma$
		versus $r=m_c/m_b$, corresponding to
		Tables~\ref{tab-br-r-dependence-jpsigamma1} (upper row)
		and~\ref{tab-br-r-dependence-jpsigamma2} (lower row).
		The gray hatched (green) bands show the LO (NLO) results,
		with combined uncertainties from the renormalization scale
		variation and the full decay widths.}
	\label{fig-r-dependence}
\end{figure}
Table~\ref{tab-br-r-dependence-jpsigamma1} and
the upper panels of Fig.~\ref{fig-r-dependence} show that the central values
for all three channels decrease monotonically with increasing $r$
at both LO and NLO. This dependence is relatively mild for $\chi_{b0}$
but more pronounced for $\chi_{b1}$ and $\chi_{b2}$. With
$m_b=4.7\,{\rm GeV}$ fixed, increasing $m_c$ from $1.3$ to
$1.7\,{\rm GeV}$ reduces the NLO branching fractions by approximately
$8\%$, $45\%$, and $52\%$ for $\chi_{b0}$, $\chi_{b1}$, and
$\chi_{b2}$, respectively, indicating a greater sensitivity to the
charm-quark mass in the latter two channels.
Table~\ref{tab-br-r-dependence-jpsigamma2} and the lower panels of Fig.~\ref{fig-r-dependence} present the branching
fractions obtained by varying $m_b$ from $4.5$ to
$4.9~\mathrm{GeV}$ at fixed $m_c=1.5~\mathrm{GeV}$.
	Increasing $m_b$ from $4.5$ to $4.9\,{\rm GeV}$ reduces the NLO
	branching fractions by approximately $49\%$, $38\%$, and $37\%$
	for $\chi_{b0}$, $\chi_{b1}$, and $\chi_{b2}$, respectively.
Within the considered ranges of $m_c$ and $m_b$,
the NLO corrections enhance the central predictions for $\chi_{b0}$
and $\chi_{b1}$ while suppressing that for $\chi_{b2}$.
The green bands in Fig.~\ref{fig-r-dependence}
	indicate smaller relative uncertainties at NLO, particularly for
	$\chi_{b2}$. This is consistent with the reduced renormalization
	scale dependence shown in
	Fig.~\ref{fig-scale-dependence-jpsigamma}.

Finally, we use the results in
Table~\ref{tab-widths-jpsigamma} to assess the prospects for
observing these decays at the LHC and $B$ factories. The production cross sections for
$\chi_{b0}$ and $\chi_{b2}$ at the LHC at $\sqrt{s}=14~{\rm TeV}$
are estimated to be $\sigma(pp\to\chi_{b0}+X)=1.5~\mu{\rm b}$
and $\sigma(pp\to\chi_{b2}+X)=2.0~\mu{\rm b}$
~\cite{Braguta:2005gw}. Taking an integrated luminosity of
$\mathcal{L}=100~{\rm fb}^{-1}$, we expect about $1.5\times10^5$
and $4.0\times10^5$ $J/\psi+\gamma$ events from $\chi_{b0}$ and
$\chi_{b2}$ decays, respectively, at the LHC. Furthermore, taking into account
$\operatorname{Br}[J/\psi\to\ell\bar\ell]\simeq12\%$
~\cite{ParticleDataGroup:2024cfk}, we expect about
$1.9\times10^4$ and $4.8\times10^4$
$\ell^+\ell^-\gamma$ events, respectively.
Although the observation prospects depend on the backgrounds
and detection efficiencies, these event yields motivate
experimental searches for the $\chi_{b0,2}\to J/\psi+\gamma$
signals at the LHC. At $B$ factories, the $\chi_{bJ}$ states
can be produced through the electromagnetic E1 transitions
$\Upsilon(2S)\to\chi_{bJ}+\gamma$. Using a sample of
$158\times10^6$ $\Upsilon(2S)$ events collected with the Belle detector~\cite{Shen:2012ei}, we estimate that about
$6\times10^6$ $\chi_{b0}$ and $10^7$ $\chi_{b1,2}$ states
were produced.
The expected numbers of $\chi_{bJ}\to J/\psi+\gamma$
decay events in this Belle sample are approximately
$6$, $160$, and $22$ for $J=0,1,2$, respectively.
The observable yields are further suppressed by the
leptonic branching fraction of the $J/\psi$ and
the detection efficiencies.
Therefore, experimental measurements of these channels remain
challenging with this Belle sample. 
Nevertheless, Belle II, with a target integrated luminosity of
$50~{\rm ab}^{-1}$~\cite{Belle-II:2025wpi}, could improve the observation prospects for
these radiative decays, particularly for the $\chi_{b1}$ channel,
provided that a substantially larger $\Upsilon(2S)$ sample is
collected.
\section{Summary}
\label{sec-summary}

In this work, we have calculated the NLO QCD corrections to
$\chi_{bJ}\to J/\psi+\gamma$ with $J=0,1,2$ within NRQCD
factorization. We have obtained all six independent helicity
amplitudes and the corresponding polarized and unpolarized
decay widths. The calculation includes the one-loop corrections to the
QED amplitude and the two-loop corrections to the QCD amplitude.
The interference between the QED and QCD amplitudes is included
in the decay widths.
For the photon fragmentation contribution, we consider both
the conventional NRQCD treatment and an alternative scheme
using the experimentally determined $J/\psi$ decay constant
\cite{Sang:2023liy}, following an approach analogous to that
employed in Ref.~\cite{Zhang:2021ted}.
In the latter scheme, the matching correction associated
with $\gamma^*\to J/\psi$ is subtracted to avoid
double counting. In the ${\rm QED}_{\rm II}$ scheme, with
$m_b=4.7~{\rm GeV}$, $m_c=1.5~{\rm GeV}$, and
$\mu_R=m_b$, the NLO corrections enhance the $\chi_{b0}$ and $\chi_{b1}$
decay widths by factors of approximately $2.8$ and $1.8$,
respectively. The NLO corrections reduce the $\chi_{b2}$ decay width
by approximately $36\%$ relative to the LO prediction.
Although the $\chi_{b0}$ and $\chi_{b1}$ decay widths
are comparable, the $\chi_{b1}$ branching fraction is
much larger because of its smaller full decay width.

The QED contribution has a markedly different impact on the
three channels. At NLO, including the QED amplitude reduces
the pure QCD prediction for $\chi_{b0}$ by approximately
$35\%$ through destructive interference, increases the
$\chi_{b2}$ decay width by approximately $20\%$, and changes
the $\chi_{b1}$ decay width by only a few percent.
Although the LO predictions for the
$\chi_{bJ}\to J/\psi+\gamma$ decay widths can differ substantially
between the two QED schemes, the unpolarized decay widths obtained in the two QED
schemes agree within approximately $2\%$ at NLO. Furthermore,
the NLO corrections also reduce
the renormalization scale dependence, especially for $\chi_{b2}$. 
However, the large NLO correction and sizable $\mu_R$-dependence in the $\chi_{b0}$ channel
indicate that higher-order corrections may still be important.
Additionally, at fixed $m_b=4.7\,{\rm GeV}$, the branching fractions decrease with increasing
$r=m_c/m_b$ at both LO and NLO, with a stronger dependence for
$\chi_{b1}$ and $\chi_{b2}$ than for $\chi_{b0}$. Increasing $m_c$ from $1.3$ to
$1.7\,{\rm GeV}$ reduces the NLO branching fractions by approximately
$8\%$, $45\%$, and $52\%$ for $\chi_{b0}$, $\chi_{b1}$,
and $\chi_{b2}$, respectively. Similarly, at fixed $m_c=1.5\,{\rm GeV}$, increasing
$m_b$ from $4.5$ to $4.9\,{\rm GeV}$ reduces the corresponding NLO
branching fractions by approximately $49\%$, $38\%$, and $37\%$,
respectively.

Finally, for $pp$ collisions at $\sqrt{s}=14\,{\rm TeV}$ and an integrated
luminosity of $100\,{\rm fb}^{-1}$, we expect about $1.9\times10^4$ and $4.8\times10^4$
$\ell^+\ell^-\gamma$ events from $\chi_{b0}$ and $\chi_{b2}$
decays, respectively. Measurements with the
Belle $\Upsilon(2S)$ sample remain challenging because of
the small expected yields. Nevertheless, a substantially
larger $\Upsilon(2S)$ sample at Belle II would improve the
observation prospects, particularly for
$\chi_{b1}\to J/\psi+\gamma$.

\section*{Acknowledgments}
The work of Y.-D. Z. and Z.-F. J. is supported by the Young and Middle-aged Scientific and Technological Innovation Team Program for Colleges and Universities of the Hubei Provincial Department of Education under Grant No. T2024029 and by the 2025 Provincial Teaching Reform Research Project of Hubei Undergraduate Universities under Grant No. 2025468.


\begin{thebibliography}{99}
	\bibitem{Bodwin:1994jh}
	G.~T.~Bodwin, E.~Braaten and G.~P.~Lepage,
	Phys. Rev. D \textbf{51}, 1125--1171 (1995)
	[erratum: Phys. Rev. D \textbf{55}, 5853 (1997)]
	doi:10.1103/PhysRevD.51.1125
	[arXiv:hep-ph/9407339 [hep-ph]].
	
	\bibitem{Belle:2010sgx}
	C.~P.~Shen \textit{et al.} [Belle],
	Phys. Rev. D \textbf{82}, 051504 (2010)
	doi:10.1103/PhysRevD.82.051504
	[arXiv:1008.1774 [hep-ex]].
	
	\bibitem{Belle:2011jbs}
	X.~L.~Wang \textit{et al.} [Belle],
	Phys. Rev. D \textbf{84}, 071107 (2011)
	doi:10.1103/PhysRevD.84.071107
	[arXiv:1108.4514 [hep-ex]].
	
	\bibitem{Belle:2019ybg}
	P.~Katrenko \textit{et al.} [Belle],
	Phys. Rev. Lett. \textbf{124}, no.12, 122001 (2020)
	doi:10.1103/PhysRevLett.124.122001
	[arXiv:1910.10915 [hep-ex]].
	
	\bibitem{Shen:2012ei}
	C.~P.~Shen \textit{et al.} [Belle],
	Phys. Rev. D \textbf{85}, 071102 (2012)
	doi:10.1103/PhysRevD.85.071102
	[arXiv:1203.0368 [hep-ex]].
	
	\bibitem{LHCb:2014ngh}
	R.~Aaij \textit{et al.} [LHCb],
	Eur. Phys. J. C \textbf{74}, no.10, 3092 (2014)
	doi:10.1140/epjc/s10052-014-3092-z
	[arXiv:1407.7734 [hep-ex]].
	
	\bibitem{Gao:2007fv}
	Y.~J.~Gao, Y.~J.~Zhang and K.~T.~Chao,
	[arXiv:hep-ph/0701009 [hep-ph]].
	
	\bibitem{Hao:2006nf}
	G.~Hao, Y.~Jia, C.~F.~Qiao and P.~Sun,
	JHEP \textbf{02}, 057 (2007)
	doi:10.1088/1126-6708/2007/02/057
	[arXiv:hep-ph/0612173 [hep-ph]].
	
	\bibitem{Shen:2020woq}
	D.~D.~Shen, C.~Y.~Lu, P.~Sun and R.~Zhu,
	[arXiv:2011.03942 [hep-ph]].
	
	\bibitem{Feng:2019zmt}
	F.~Feng, Y.~Jia, Z.~Mo, W.~L.~Sang and J.~Y.~Zhang,
	Phys. Lett. B \textbf{850} (2024), 138506
	doi:10.1016/j.physletb.2024.138506
	[arXiv:1901.08447 [hep-ph]].
	
	\bibitem{Huang:2022dfw}
	X.~D.~Huang, B.~Gong and J.~X.~Wang,
	JHEP \textbf{02}, 049 (2023)
	doi:10.1007/JHEP02(2023)049
	[arXiv:2212.03631 [hep-ph]].
	
	\bibitem{Sang:2022kub}
	W.~L.~Sang, F.~Feng, Y.~Jia, Z.~Mo and J.~Y.~Zhang,
	Phys. Lett. B \textbf{843}, 138057 (2023)
	doi:10.1016/j.physletb.2023.138057
	[arXiv:2202.11615 [hep-ph]].
	
	\bibitem{Zhang:2021ted}
	Y.~D.~Zhang, F.~Feng, W.~L.~Sang and H.~F.~Zhang,
	JHEP \textbf{12}, 189 (2021)
	doi:10.1007/JHEP12(2021)189
	[arXiv:2109.15223 [hep-ph]].
	
	\bibitem{Zhang:2022nuf}
	Y.~D.~Zhang, W.~L.~Sang and H.~F.~Zhang,
	Phys. Rev. Lett. \textbf{129}, no.11, 112002 (2022)
	doi:10.1103/PhysRevLett.129.112002
	[arXiv:2205.06124 [hep-ph]].
	
	\bibitem{Belle:2014wam}
	S.~D.~Yang \textit{et al.} [Belle],
	Phys. Rev. D \textbf{90}, no.11, 112008 (2014)
	doi:10.1103/PhysRevD.90.112008
	[arXiv:1409.7644 [hep-ex]].
	
	\bibitem{Zhang:2023mky}
	Y.~D.~Zhang, X.~W.~Bai, F.~Feng, W.~L.~Sang and M.~Z.~Zhou,
	Phys. Rev. D \textbf{108}, no.11, 114030 (2023)
	doi:10.1103/PhysRevD.108.114030
	[arXiv:2310.07453 [hep-ph]].
	
	\bibitem{Yang:2026kvh}
	H.~Yang, L.~Zuo and B.~Long,
	Phys. Rev. D \textbf{113}, no.11, 116015 (2026)
	doi:10.1103/7bbf-37y2
	[arXiv:2603.14847 [hep-ph]].
	
	\bibitem{Sang:2023liy}
	W.~L.~Sang, F.~Feng, Y.~Jia, Z.~Mo, J.~Pan and J.~Y.~Zhang,
	Phys. Rev. Lett. \textbf{131}, no.16, 161904 (2023)
	doi:10.1103/PhysRevLett.131.161904
	[arXiv:2306.11538 [hep-ph]].
	
	\bibitem{Haber:1994pe}
	H.~E.~Haber,
	[arXiv:hep-ph/9405376 [hep-ph]].
	
	\bibitem{Jacob:1959at}
	M.~Jacob and G.~C.~Wick,
	Annals Phys. \textbf{7}, 404-428 (1959)
	doi:10.1016/0003-4916(59)90051-X

	\bibitem{Petrelli:1997ge}
	A.~Petrelli, M.~Cacciari, M.~Greco, F.~Maltoni and M.~L.~Mangano,
	Nucl. Phys. B \textbf{514} (1998), 245-309
	doi:10.1016/S0550-3213(97)00801-8
	[arXiv:hep-ph/9707223 [hep-ph]].
	
	\bibitem{Bodwin:2002cfe}
	G.~T.~Bodwin and A.~Petrelli,
	Phys. Rev. D \textbf{66} (2002), 094011
	[erratum: Phys. Rev. D \textbf{87} (2013) no.3, 039902]
	doi:10.1103/PhysRevD.66.094011
	[arXiv:hep-ph/0205210 [hep-ph]].
	
	\bibitem{Beneke:1997jm}
	M.~Beneke, A.~Signer and V.~A.~Smirnov,
	Phys. Rev. Lett. \textbf{80} (1998), 2535-2538
	doi:10.1103/PhysRevLett.80.2535
	[arXiv:hep-ph/9712302 [hep-ph]].
	
	\bibitem{Czarnecki:1997vz}
	A.~Czarnecki and K.~Melnikov,
	Phys. Rev. Lett. \textbf{80} (1998), 2531-2534
	doi:10.1103/PhysRevLett.80.2531
	[arXiv:hep-ph/9712222 [hep-ph]].
	
	\bibitem{Marquard:2014pea}
	P.~Marquard, J.~H.~Piclum, D.~Seidel and M.~Steinhauser,
	Phys. Rev. D \textbf{89} (2014) no.3, 034027
	doi:10.1103/PhysRevD.89.034027
	[arXiv:1401.3004 [hep-ph]].
	
	\bibitem{Binosi:2003yf}
	D.~Binosi and L.~Theu{\ss}l,
	Comput. Phys. Commun. \textbf{161} (2004), 76-86
	doi:10.1016/j.cpc.2004.05.001
	[arXiv:hep-ph/0309015 [hep-ph]].
	
	\bibitem{Jia:2024ini}
	Y.~Jia, Z.~Mo and J.~Y.~Zhang,
	[arXiv:2408.17448 [hep-ph]].
	
	\bibitem{Huang:2023pmn}
	X.~D.~Huang, B.~Gong, R.~C.~Niu, H.~M.~Yu and J.~X.~Wang,
	JHEP \textbf{02}, 055 (2024)
	doi:10.1007/JHEP02(2024)055
	[arXiv:2311.04751 [hep-ph]].
	
	\bibitem{Zeng:2026ois}
	C.~Zeng, B.~Gong, J.~X.~Wang, R.~Niu, X.~D.~Huang and C.~Li,
	[arXiv:2603.26199 [hep-ph]].
	
	\bibitem{Hahn:2000kx}
	T.~Hahn,
	Comput. Phys. Commun. \textbf{140}, 418-431 (2001)
	doi:10.1016/S0010-4655(01)00290-9
	[arXiv:hep-ph/0012260 [hep-ph]].
	
	\bibitem{Shtabovenko:2016sxi}
	V.~Shtabovenko, R.~Mertig and F.~Orellana,
	Comput. Phys. Commun. \textbf{207}, 432-444 (2016)
	doi:10.1016/j.cpc.2016.06.008
	[arXiv:1601.01167 [hep-ph]].
	
	\bibitem{Feng:2012tk}
	F.~Feng and R.~Mertig,
	[arXiv:1212.3522 [hep-ph]].
	
	\bibitem{Beneke:1997zp}
	M.~Beneke and V.~A.~Smirnov,
	Nucl. Phys. B \textbf{522}, 321-344 (1998)
	doi:10.1016/S0550-3213(98)00138-2
	[arXiv:hep-ph/9711391 [hep-ph]].
	
	\bibitem{Feng:2012iq}
	F.~Feng,
	Comput. Phys. Commun. \textbf{183}, 2158-2164 (2012)
	doi:10.1016/j.cpc.2012.03.025
	[arXiv:1204.2314 [hep-ph]].
	
	\bibitem{Smirnov:2014hma}
	A.~V.~Smirnov,
	Comput. Phys. Commun. \textbf{189}, 182-191 (2015)
	doi:10.1016/j.cpc.2014.11.024
	[arXiv:1408.2372 [hep-ph]].
	
	\bibitem{Patel:2015tea}
	H.~H.~Patel,
	Comput. Phys. Commun. \textbf{197}, 276-290 (2015)
	doi:10.1016/j.cpc.2015.08.017
	[arXiv:1503.01469 [hep-ph]].
	
	\bibitem{Liu:2017jxz}
	X.~Liu, Y.~Q.~Ma and C.~Y.~Wang,
	Phys. Lett. B \textbf{779}, 353-357 (2018)
	doi:10.1016/j.physletb.2018.02.026
	[arXiv:1711.09572 [hep-ph]].
	
	\bibitem{Liu:2022mfb}
	Z.~F.~Liu and Y.~Q.~Ma,
	Phys. Rev. Lett. \textbf{129}, no.22, 222001 (2022)
	doi:10.1103/PhysRevLett.129.222001
	[arXiv:2201.11637 [hep-ph]].
	
	\bibitem{Liu:2021wks}
	X.~Liu and Y.~Q.~Ma,
	Phys. Rev. D \textbf{105} (2022) no.5, L051503
	doi:10.1103/PhysRevD.105.L051503
	[arXiv:2107.01864 [hep-ph]].
	
	\bibitem{Liu:2022chg}
	X.~Liu and Y.~Q.~Ma,
	Comput. Phys. Commun. \textbf{283}, 108565 (2023)
	doi:10.1016/j.cpc.2022.108565
	[arXiv:2201.11669 [hep-ph]].
	
	\bibitem{ParticleDataGroup:2024cfk}
	S.~Navas \textit{et al.} [Particle Data Group],
	Phys. Rev. D \textbf{110}, 030001 (2024)
	doi:10.1103/PhysRevD.110.030001.
	
	\bibitem{Chetyrkin:2000yt}
	K.~G.~Chetyrkin, J.~H.~K{\"u}hn and M.~Steinhauser,
	Comput. Phys. Commun. \textbf{133} (2000), 43-65
	doi:10.1016/S0010-4655(00)00155-7
	[arXiv:hep-ph/0004189 [hep-ph]].
	
	\bibitem{Sang:2009jc}
	W.~L.~Sang and Y.~Q.~Chen,
	Phys. Rev. D \textbf{81} (2010), 034028
	doi:10.1103/PhysRevD.81.034028
	[arXiv:0910.4071 [hep-ph]].
	
	\bibitem{Fan:2012dy}
	Y.~Fan, J.~Lee and C.~Yu,
	Phys. Rev. D \textbf{87} (2013) no.9, 094032
	doi:10.1103/PhysRevD.87.094032
	[arXiv:1211.4111 [hep-ph]].
	
	\bibitem{Eichten:1995ch}
	E.~J.~Eichten and C.~Quigg,
	Phys. Rev. D \textbf{52}, 1726-1728 (1995)
	doi:10.1103/PhysRevD.52.1726
	[arXiv:hep-ph/9503356 [hep-ph]].
	
	\bibitem{Chung:2021efj}
	H.~S.~Chung,
	JHEP \textbf{09}, 195 (2021)
	doi:10.1007/JHEP09(2021)195
	[arXiv:2106.15514 [hep-ph]].
	
	\bibitem{Braguta:2005gw}
	V.~V.~Braguta, A.~K.~Likhoded and A.~V.~Luchinsky,
	Phys. Rev. D \textbf{72} (2005), 094018
	doi:10.1103/PhysRevD.72.094018
	[arXiv:hep-ph/0506009 [hep-ph]].
	
	\bibitem{Belle-II:2025wpi}
	F.~Bernlochner \textit{et al.} [Belle-II],
	[arXiv:2503.24155 [hep-ex]].
	
\end{thebibliography}
\end{document}